\documentclass[journal]{IEEEtran}
\usepackage{xurl} 
\usepackage{hyperref}
\usepackage{amsmath,amsfonts}
\usepackage{array}
\usepackage[noadjust]{cite}
\usepackage{textcomp}
\usepackage{stfloats}
\usepackage{url}
\usepackage{verbatim}
\usepackage{graphicx}
\def\BibTeX{{\rm B\kern-.05em{\sc i\kern-.025em b}\kern-.08em
    T\kern-.1667em\lower.7ex\hbox{E}\kern-.125emX}}
\usepackage{balance}
\usepackage{tikz}
\usetikzlibrary{fadings}
\usetikzlibrary{patterns}
\usetikzlibrary{shadows.blur}
\usetikzlibrary{shapes}   
\usepackage{siunitx}
\DeclareSIUnit\mt{\milli\tesla} 
\usepackage[utf8]{inputenc}
\usepackage[english]{babel}
\usepackage{xurl} 
\usepackage[linesnumbered,ruled,vlined]{algorithm2e}
\usepackage{array}
\usepackage[noadjust]{cite}
\usepackage{textcomp}
\usepackage{url}
\usepackage{verbatim}
\usepackage{graphicx}
\def\BibTeX{{\rm B\kern-.05em{\sc i\kern-.025em b}\kern-.08em
    T\kern-.1667em\lower.7ex\hbox{E}\kern-.125emX}}
\usepackage{balance}
\usepackage{multirow} 
\usepackage{setspace}
\usepackage{graphicx}
\usepackage{url}
\usepackage{subcaption}
\usepackage{amsmath}
\usepackage{algcompatible}
\usepackage{array}
\usepackage{amsthm}
\usepackage{adjustbox}
\usepackage{txfonts}
\usepackage{lipsum}  
\usepackage{color}

\usepackage{pgfplots}
\usepackage{tikz}
\usetikzlibrary{fadings}
\usetikzlibrary{patterns}
\usetikzlibrary{shadows.blur}
\usetikzlibrary{shapes} 
\usepackage{algpseudocode}

\let\oldnl\nl
\newcommand{\nonl}{\renewcommand{\nl}{\let\nl\oldnl}}
\newcolumntype{L}[1]{>{\raggedright\let\newline\\\arraybackslash\hspace{0pt}}m{#1}}
\newcolumntype{C}[1]{>{\centering\let\newline\\\arraybackslash\hspace{0pt}}m{#1}}
\newcolumntype{R}[1]{>{\raggedleft\let\newline\\\arraybackslash\hspace{0pt}}m{#1}}

\algnewcommand\KwEvl{\textbf{Evaluation:}}
\usepackage{booktabs}

\usepackage{makecell}
\usepackage{amsmath}
\usepackage{siunitx}
\usepackage{caption} 

\usepackage{pgfplots}
\pgfplotsset{compat=1.18}

\begin{document}
\title{Explainable AI-assisted Resource Allocation in 6G V2X Communications}
\title{Explainable AI-aided Feature Selection and Model Reduction for DRL-based V2X Resource Allocation}

\author{Nasir~Khan,~\IEEEmembership{Graduate Student Member,~IEEE}, Asmaa~Abdallah,~\IEEEmembership{Member,~IEEE},\\Abdulkadir~Celik,~\IEEEmembership{Senior~Member,~IEEE},~Ahmed~M.~Eltawil,~\IEEEmembership{Senior Member,~IEEE},\\ and~Sinem Coleri,~\IEEEmembership{Fellow,~IEEE}

\thanks{Nasir~Khan and Sinem Coleri are with the department of Electrical and Electronics Engineering, Koc University, Istanbul, Turkey, email: $\lbrace$nkhan20, scoleri$\rbrace$@ku.edu.tr. This work is supported by Scientific and Technological Research Council of Turkey Grant $\#$119C058 and Ford Otosan.}
 \thanks{Asmaa Abdallah, Abdulkadir Celik, and Ahmed M. Eltawil are with the Computer, Electrical, and Mathematical Sciences and Engineering Division, King Abdullah University of Science and Technology, Thuwal 23955, Saudi Arabia (e-mail:  asmaa.abdallah@kaust.edu.sa; abdulkadir.celik@kaust.edu.sa;  ahmed.eltawil@kaust.edu.sa)}}
\maketitle
\begin{abstract}  
Artificial intelligence (AI) is expected to significantly enhance radio resource management (RRM) in sixth-generation (6G) networks. However, the lack of explainability in complex deep learning (DL) models poses a challenge for practical implementation. This paper proposes a novel explainable AI (XAI)-based framework for feature selection and model complexity reduction in a model-agnostic manner. Applied to a multi-agent deep reinforcement learning (MADRL) setting, our approach addresses the joint sub-band assignment and power allocation problem in cellular vehicle-to-everything (V2X) communications. We propose a novel two-stage systematic explainability framework leveraging feature relevance-oriented XAI to simplify the DRL agents. While the former stage generates a state feature importance ranking of the trained models using Shapley additive explanations (SHAP)-based importance scores, the latter stage exploits these importance-based rankings to simplify the state space of the agents by removing the least important features from the model's input. Simulation results demonstrate that the XAI-assisted methodology achieves $\sim$97\% of the original MADRL sum-rate performance while reducing optimal state features by $\sim$28\%, average training time by $\sim$11\%, and trainable weight parameters by $\sim$46\% in a network with eight vehicular pairs.
\end{abstract}

\IEEEpeerreviewmaketitle

\section{Introduction}
\noindent \IEEEPARstart{C}{ellular} vehicle-to-everything (C-V2X) networks have emerged as key enablers for intelligent transportation systems, leveraging existing cellular infrastructure 
to improve road safety, traffic efficiency, and in-vehicle entertainment. C-V2X technology promises widespread coverage, high reliability, and efficient spectrum use, even in high-mobility scenarios \cite{5G-V2X}. Advanced C-V2X applications demand transmission latencies of a few milliseconds and 99.999\% reliability \cite{tutorial}. 
As vehicular networks expand and radio resource management becomes increasingly complex, existing centralized resource allocation approaches in cellular networks struggle to meet diverse quality-of-service (QoS) requirements, particularly ultra-reliable and low-latency needs.


The recent decade has witnessed a surge in publications unveiling the merits of artificial intelligence (AI) and deep learning (DL) methods, suggesting their potential to mitigate the aforementioned challenges of model-based approaches \cite{celik2024GenAI}. In particular,  deep reinforcement learning (DRL) has been shown effective in solving hard-to-optimize non-convex combinatorial problems
efficiently and has widely been adopted to deal with high-dimensional state-action spaces of vehicular communications
applications \cite{DRL23_0}. Nevertheless, DRL-based resource allocation relies on complex deep neural networks (DNNs)
with multiple layers, making them difficult to understand due to their black-box nature and lack of explainability \cite{our_work}. This complexity obscures the influence of input state features on the model's output and the logical reasoning behind decisions.  This lack of explainability hinders the adoption of DRL-based resource allocation in safety-critical vehicular applications, as the decision-making processes are not easily understandable.

\subsection{Relevant works}
Resource sharing in C-V2X applications requires a judicious allocation for mitigating interference and optimizing resource utilization for efficient
 vehicular communications. In centralized resource management in C-V2X, the BS collects
the channel statistics to monitor the quality of each link and formulate
the optimal resource allocation objective function \cite{graph_solution, matching, graph_solution0, uncertain_CSI}. 
For instance,  centralized graph theory-based proposals \cite{ graph_solution}  and hierarchical
optimization theory-based solutions \cite{matching}  require executing iterative algorithms at the central controller with up-to-date global channel state information (CSI) availability. To alleviate the  CSI acquisition overhead, low-complexity algorithms based on 
only large-scale fading information \cite{graph_solution0} or imperfect CSI assumptions \cite{uncertain_CSI} have been proposed.  However, the V2X resource allocation problem is
often modeled as a combinatorial optimization problem with
nonlinear constraints, which are difficult to solve using traditional optimization algorithms in real time. Further, these works \cite{graph_solution, matching, graph_solution0, uncertain_CSI} consider Shannon rate-based communication, ignoring the reliability aspect of V2V links in terms of the decoding error probability, implicitly assuming infinite blocklength availability, making them unsuitable for the low-latency, high-reliability requirements of V2X scenarios addressed in this study.

Many existing studies have considered different DRL-based solutions for solving resource allocation problems in V2X communication networks \cite{DRL23_1, DRL23_7, QOS, LSTM }. Progress has also been made to integrate DRL with federated learning \cite{DRL23_3}, \cite{DRL23_5}, meta-learning \cite{MetaV2X}, graph theory \cite{graphDRL} and digital twin-driven V2X networks \cite{DRL24_0} for improved performance and fast adaptability of the resource allocation policy in the dynamic V2X environments. However, to the best of the authors' knowledge, none have addressed the open problem of providing explainability to these solutions. These aforementioned studies  \cite{DRL23_1, DRL23_3, DRL23_5,  DRL23_7, MetaV2X, graphDRL, QOS, LSTM, DRL24_0 } commonly focus on the performance and robustness of the underlying DRL-based strategies and can yield a good solution for V2X resource allocation problems. However, they do not provide insights into the workings of these complex model-based solutions and, as such, suffer from a performance-explainability trade-off.

To strike a good balance between performance and explainability trade-off, explainable AI (XAI) is gaining significant momentum in wireless communications to aid
the understanding of complex AI models \cite{trustworthy}. XAI encompasses processes and methods that reveal the inner workings of complex AI models, helping to balance the trade-off between explainability and performance inherent to these models. The Shapley additive explanations (SHAP) method is particularly effective for feature selection, as it assigns importance values to each feature based on their contribution to the model’s predictions, allowing for the exclusion of less significant features without sacrificing performance \cite{XAI_wireless}. SHAP has a strong theoretical foundation based on Shapley values in game theory and exhibits several desirable properties compared to other XAI methods. First, SHAP solutions satisfy three essential properties for any additive feature attribution model: local accuracy, missingness, and consistency \cite{lundberg}. Second, SHAP combines the local interpretable model-agnostic explanations (LIME) \cite{lime} method, which explains individual predictions using an interpretable model, with Shapley values. This integration allows for faster computation of explanations for complex learning models than directly calculating Shapley values.

Within the wireless communications literature, research on the XAI  application remains limited. Several studies have begun integrating XAI techniques to enhance transparency and understanding within complex network systems. For instance, classical XAI methods and SHAP have been recommended to analyze the root cause of Service Level Agreement (SLA) violation prediction in a 5G network slicing setup \cite{XAI1}.  Kernel-SHAP is invoked to quantify the importance of input feature contributions to the model's outcome and demystify model behavior for short-term resource reservation (STRR)  in network slicing operations \cite{XAI3}. A two-stage pipeline is introduced for robust network intrusion detection using extreme gradient boosting (XGBoost), and SHAP is employed to explain intrusion attacks \cite{XAI4}. Further, XAI is integrated with federated learning to predict and interpret network slices' latency key performance indicator (KPI) \cite{XAI5}.  XAI is utilized to design a composite reward mechanism assisted by SHAP importance scores to encourage DRL agents to learn the best actions for specific network slices in SLA-aware 6G network slicing setup \cite{XAI_DRL }. A trust-aware, explainable federated DRL model is developed using SHAP values to compute the trust score for trajectory and velocity decision-making in autonomous driving. The trust score is used to select and aggregate vehicles participating in the federated learning process \cite{AVs}. The scope of the above works \cite{XAI3, XAI4, XAI5, XAI_DRL, AVs} does not extend beyond providing insights into the feature relevance in the decision-making process of the underlying machine learning models in different wireless settings. Additionally, the XAI metrics (trust score, XRL reward) in  \cite{XAI_DRL, AVs} are tied to the underlying algorithm and cannot be applied directly to other learning algorithms without modification.

XAI-based methods for feature selection have garnered significant attention in wireless communications to determine the primary attributes influencing the model decision and enhancing model performance. Feature selection aims to identify a minimal set of features that optimizes the objective function without significantly degrading performance. Traditional filter-based feature selection methods assess importance based on data characteristics but overlook algorithm dependence and feature interactions. Wrapper methods (e.g., Recursive Feature Elimination) improve accuracy but are computationally expensive for large datasets. XAI-based feature selection methods are generally architecture-specific, data-dependent, and provide local explanations \cite{lime}. In contrast, SHAP stands out as a popular and stable XAI method owing to its robust mathematical foundation embedded in the game theory concept of Shapley values. It provides a
fair allocation of rewards among inputs, considering their individual/collective contributions and providing both local/global explanations.

To our knowledge, no prior work has provided a systematic approach for generating feature importance explanations for complex AI models in multi-agent deep reinforcement learning (MADRL) settings or applied these explanations to enhance understanding or reduce the execution time and model complexity (e.g., via feature selection). However, similar methods have been used for feature importance in other networking areas.  For instance,  feature importance scoring for tree-based classifiers in wireless intrusion detection using the SHAP XAI method is derived in \cite{bhandari2020feature}. By leveraging these importance scores, the AI classifiers are simplified, resulting in a 90\% reduction in model size without any significant performance loss. XAI-based feature selection methodology comparing counterfactual and rule-based explanation scores with SHAP scores for feature ranking is used to simplify trained models in energy-efficient resource management in 5G networks \cite{XAI_ICC}.

\subsection{Contributions}
In this work, we present a distributed multi-agent deep reinforcement learning (MADRL) algorithm to jointly optimize transmit power and spectrum sub-band assignment, enhancing the performance of vehicle-to-network (V2N) and vehicle-to-vehicle (V2V) links in ultra-reliable and
low-latency communications (URLLC)-enabled V2X communications network. Utilizing the SHAP method from XAI, we design a post-hoc and model-agnostic explainability pipeline that improves understanding of DRL agent inferences and simplifies the agent's inputs through feature selection based on importance rankings.\footnote{A preliminary version of this work is described as a case study in \cite{our_work}, where we devise a fixed threshold-based input feature selection
strategy utilizing the SHAP-based feature importance ranking in a power control problem in V2X networks. Different from our previous work \cite{our_work}, we extend the framework to include sub-band allocation and propose a novel post-hoc explainability framework with an automated variable state feature selection strategy.} The  contributions of this paper are summarized as follows:

\begin{itemize}
    \item We formulate the joint optimization problem of spectrum and transmit power allocation with the objective of maximizing the sum-rate of V2N and V2V links under reliability, latency, and transmit power constraints in a URLLC-enabled vehicular network. We propose a MADRL algorithm with centralized learning and decentralized execution to solve the formulated optimization problem. Specifically,   multiple DQNs are distributively executed at the V2V transmitters, whereas the weight parameters are centrally trained at the base station (BS) using a common reward function for all agents to ease implementation and improve stability.
    
    \item We propose a novel two-stage systematic explainability framework leveraging feature relevance-oriented XAI to simplify the DRL agents. The former stage involves generating state feature importance ranking of the trained models using the SHAP-based importance scores. The latter stage exploits these importance-based rankings to simplify the state space of the agents, whereby the least important features are removed from the model's input, i.e., by masking feature values and observing the effect on the model's performance using an automated novel variable feature selection method.

    \item We quantitatively validate the effectiveness of the proposed XAI-based methodology. Via extensive simulations, we demonstrate that our proposed methodology can simplify the model in terms of the average training time, the number of broadcast parameters, and the number of optimal state features required for training for different network configurations without significant performance loss.
\end{itemize}

\subsection{Notations and Paper Organization}

The boldface lowercase letter is used to represent a column vector. 
 $\leftarrow$ denotes the assignment operation. $\|\mathbf{x}\|$ denotes the length of  $\mathbf{x}$. $|\cdot|$ is the absolute function operator and $f \equiv g$ denotes that function $f$ is equivalent to $g$.  Key mathematical notations are summarized in Table \ref{table:notations}.

The rest of the paper is organized as follows. Section \ref{sec:system} describes the vehicular URLLC system model and assumptions used in the paper. Section \ref{sec:opt_prob} presents the mathematical formulation of the joint transmit power and spectrum sub-band allocation in vehicular URLLC system. Section \ref{MARL-solution} provides a brief background on the design of the DRL framework and describes the proposed multi-agent DRL-based algorithm for the transmit power and spectrum sub-band allocation problem. Section \ref{XAI} explains the proposed systematic  SHAP-based XAI methodology for generating the feature importance ranking and describes the proposed feature selection algorithm.  Section \ref{sec:simulation} evaluates the performance of the proposed solution strategy. Finally, conclusions and future research directions are presented in Section \ref{sec:conclusion}.

 \begin{table}[!t]
    \centering
    \footnotesize
    \caption{\textsc{Table of notations used in this work}}\footnotesize
         \begin{tabular}{C{1.7cm}|L{5.7cm}}\label{table:notations}
        \textbf{Notation} & \textbf{Stands For}\\
         \hline
        $\mathcal{N}, \mathcal{K}$ & Set of V2N links and V2V links\\
        \hline
        ${g}^{(t)}_{k}[n]$ &  Direct channel gain of the $k$-th V2V link \\
         \hline
        $g^{(t)}_{k^{\prime} \rightarrow k}[n]$ & Interfering channel gain from $k^{\prime}$-th V2V link to $k$-th V2V receiver\\
        \hline
        ${g}^{(t)}_{k \rightarrow B}[n]$ & Interfering channel gain from $k$-th V2V link to to BS\\
        \hline
        ${ g}^{(t)}_{n \rightarrow B}[n]$ &  Direct channel gain from $n$-th V2N link  to the BS\\ 
        \hline
        ${ g}^{(t)}_{n \rightarrow k}[n]$ & Interfering channel gain from $n$-th V2N link to $k$-th V2V receiver\\
        \hline
        \rule{0pt}{2.5ex} 
         $P_{n}^{V2N\,(t)}$ &  Transmit power of the $n$-th V2N link\\
         \hline 
         \rule{0pt}{2.5ex} 
         $ P_{k}^{V2V\,(t)}[n]$ &  Transmit power of the  $k$-th V2V link\\
         \hline 
         \rule{0pt}{2.5ex} 
         $\eta^{(t)}_{k}[n]$ & Resource allocation indicator of $k$-th V2V pair \\
        \hline
        $\varepsilon^{(t)}_{k}[n]$  & decoding error probability at the $k$-th V2V receiver\\
        \hline
        $\epsilon_{\max, k}$  &  Maximum  error probability \\
        \hline
        $\mathcal{D}, D$ & Replay buffer and mini-batch size\\
        \hline 
        \rule{0pt}{2.5ex} 
        $\mathcal{S}_{k}$  & SHAP matrix for  $k$-th agent\\ 
        \hline
        $\mathcal{X}_{BG}, \mathcal{X}$ & Background and hold-out datasets\\
        \hline
        $\mathbf{x}$ & Input state feature vector\\
        \hline
        $ f_k(\mathbf{x})$ & Prediction model for $k$-th agent\\
        \hline
        \rule{0pt}{2.5ex} 
         $\boldsymbol{\phi}_k(\mathbf{x})$ & SHAP\_values for the $k$-th agent\\
        \hline
        \rule{0pt}{2.5ex}
        $\mathbf{\Tilde{\Phi}}_{k,l}$ &  Transformed SHAP\_values  for the $k$-th  agent \\
        \hline \rule{0pt}{2.5ex} 
        $\mathbf{\widehat{{\Phi}}}$  & Average $\mathbf{\Tilde{\Phi}}_{k,l}$ across the $K$ agents \\
        \hline 
        $L, M$ & Total number of state features / valid actions\\
        \hline
        $\Delta$ & Precision threshold\\
        \hline
        $\alpha$ & Average network performance metric\\
        \hline
        \end{tabular}
\end{table}

\section{System Model and Problem Formulation} \label{sec:system}
This section first provides the system model of the C-V2X communications network. Then, it formulates a resource allocation problem that aims to maximize the sum-rate for the V2N links as well as to ensure reliable information transmission for the V2V links.

\subsection{System Model}
\begin{figure}[t]
 \centering
\includegraphics[width= 0.8\linewidth]{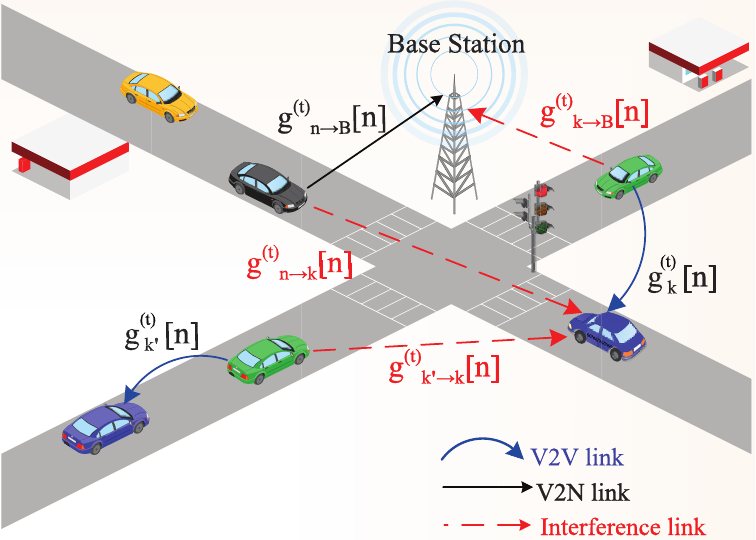}
\caption{System model for the V2X communication network.} \label{fig:V2V}
\end{figure}

We consider a single-cell C-V2X communications network covered by a Road-Side Unit (RSU) acting as a single-antenna BS, as shown in Fig. \ref{fig:V2V}. Each vehicle is equipped with a single-antenna On-Board Unit (OBU), supporting high data rate uplink services and reliable safety message sharing. C-V2X includes two operation modes to support diverse vehicular applications: V2N and V2V mode. V2N links support high data rate applications (e.g., video streaming, location tracking, map updates) via cellular interfaces using the Uu interface of 5G new radio (NR) \cite{tutorial}. V2V links transmit safety-related information through device-to-device (D2D) communications, requiring low latency and high reliability. We represent the sets of V2N and V2V links as $\mathcal{N}$ and $\mathcal{K}$, with cardinalities $N$ and $K$, respectively. Each V2N link occupies a unique sub-band with fixed transmission power. In cases where $N > K$, we introduce $N - K$ virtual V2V connections. Our focus is on Mode 4 in the cellular V2X architecture, where vehicles autonomously select radio resources for V2V communications \cite{R11}. Each V2V link can reuse an orthogonal sub-band of the V2N links to improve spectrum utilization, with multiple V2V links potentially reusing the same sub-band, which may cause interference.

\subsubsection{Channel Models}
Consider a slotted communication system with scheduling time slots indexed by $t$, where each slot has a duration of   $\tau$. We consider a quasi-static block fading channel model where orthogonal frequency division multiplexing (OFDM) technology is utilized to group consecutive subcarriers to form a spectrum sub-band. The  channel coherence time is given by ${T}_c=\sqrt{\frac{9}{16 \pi f_D^2}},$ where $f_D = \frac{f_c \cdot v_s}{c}$ is  the Doppler frequency, $f_c$ is the carrier frequency, $v_s$ is the vehicle velocity, and $c$ is the speed of light  \cite{channel_coherence}. ${T}_c > \tau$ holds for moderate vehicular speeds, and the channel state information (CSI) can be regarded as constant throughout the slot duration.
Then, the direct channel gain of the $k$-th V2V link over the $n$-th sub-band in coherence time slot  $t$ is expressed as $ g^{(t)}_{k}[n]=\alpha^{(t)}_{k} \left| {h}^{(t)}_{k}[n]\right|^2, $ where ${h}^{(t)}_{k}[n] \sim \mathcal{C N}(0,1)$ is a complex Gaussian variable representing the Rayleigh fading, and $\alpha_{k}^{(t)}$ captures the large-scale fading effect, including path loss and shadowing, which is typically frequency-independent and changes slowly since it primarily depends on the locations of the transmitter and receiver of the V2V pair. The  channel gain between  the $n$-th V2N transmitter and the BS over the $n$-th sub-band at slot $t$ is defined as ${ g}^{(t)}_{n \rightarrow B}[n]=\alpha_{n\rightarrow B}^{(t)} \left|h_{n\rightarrow B}^{(t)}[n]\right|^2$, where subscript $n\rightarrow B$ denotes the $n$-th V2N link traversing the BS.

In the case of $n$-th sub-band's reuse, the interfering channel gains
between the $i$-th and the $j$-th V2V/V2N links are defined as $g_{i\rightarrow j}^{(t)}[n]=\alpha_{i\rightarrow j}^{(t)} \left|h_{i\rightarrow j}^{(t)}[n]\right|^2$, where $i\neq j$, $i\in \{k,n\}$, $j \in \{k,B\}$, $\alpha_{i\rightarrow j}^{(t)}$ represents the large-scale fading, and $ h_{i\rightarrow j}^{(t)}[n]$ denotes the   small-scale fading. 


\subsubsection{Achievable Rates}
  
The achievable data rate of the $n$-th V2N link at time slot $t$ is expressed as
\begin{equation} \label{V2N_rate}
R_n^{V2N\,(t)}[n]=B_w \log _2\left(1+\gamma_n^{V2N\,(t)}[n]\right),
\end{equation}
where $B_w$ indicates the channel bandwidth occupied by each V2N link and $\gamma_n^{V2N\,(t)}[n]$ is the  received signal-to-interference-plus-noise ratio (SINR)  of the $n$-th V2N link over the $n$-th sub-band at time slot $t$, which is defined as
\begin{equation}
\gamma_n^{V2N\,(t)}[n]=\frac{P_n^{V2N\,(t)} {g}^{(t)}_{n \rightarrow B}[n] }{\sigma^2+\displaystyle\sum_{k \in \mathcal{K}} \eta^{(t)}_{k}[n] P^{V2V\,(t)}_{k}[n] g^{(t)}_{k \rightarrow B}[n]},
\end{equation}
where  $P_n^{V2N\,(t)}$ and $P_{k}^{V2V\,(t)}[n]$ denote the  transmit powers of the $n$-th V2N link and the $k$-th V2V link over the $n$-th sub-band, respectively; $\sigma^2$ represents the variance of additive white Gaussian noise; and  $\eta^{(t)}_{k}[n] \in\{0,1\}$ is the resource allocation indicator with $\eta^{(t)}_{k}[n]=1$ implying the  $k$-th V2V link uses the spectrum of the $n$-th V2N link and $\eta^{(t)}_{k}[n]=0$ otherwise.  We assume that each V2V link can only use one sub-band at the same time, i.e., $\sum_{n \in \mathcal{N}} \eta^{(t)}_{k}[n]=1$, while the orthogonal sub-band of each V2N link can be reused by multiple V2V pairs.

 URLLC-enabled  V2V communications are mainly responsible for the reliable dissemination of safety-critical messages using short code block lengths, rendering the classic Shannon capacity no longer appropriate to describe the maximum achievable data rate. In the context of finite block length regime,  the maximum achievable  information rate   of the $k$-th V2V link at time slot $t$ over the $n$-th sub-band,  which can be decoded with block error probability no greater than $\varepsilon_{k}$  is given by the  normal approximation \cite{R13}
\begin{equation} \label{v2v_rate}
R_{k}^{V2V\,(t)}[n] =B_w\left(C\left(\gamma_{k}^{V2V\,(t)}[n] \right)-\sqrt{\frac{V^{(t)}_{k}[n]}{m} } \frac{f_{Q^{-1}}(\varepsilon^{(t)}_{k}[n])}{\ln 2}\right),
\end{equation}
 where $C\left(\gamma_{k}^{V2V\,(t)}[n]\right)=log _{2}\left(1+\gamma_{k}^{V2V\,(t)}[n]\right)$ is the Shannon rate, $V^{(t)}_k[n] \triangleq 1-\left(1+\gamma_{k}^{V2V\,(t)}[n]\right)^{-2}$ is the channel dispersion, $m$ is the blocklength allocated to the $k$-th V2V link, $\varepsilon^{(t)}_{k}[n]\in(0,1)$ is the  decoding error probability at the $k$-th V2V receiver over $n$-th sub-band,   $f_{Q^{-1}}(\cdot)$ is the inverse Gaussian tail function $f_{Q}\left(x\right)= \frac{1}{\sqrt{2 \pi}} \int_{x}^{\infty} e^{-t^2 / 2} d t$, and $\gamma_{k}^{V2V\,(t)}[n]$ is the SINR of the $k$-th V2V link over the $n$-th sub-band at time slot $t$ defined as 
 
\begin{equation}
\gamma_{k}^{V2V\,(t)}[n]=\frac{P_{k}^{V2V\,(t)}[n] {g}^{(t)}_{k}[n] }{\sigma^2+I^{(t)}_{k}[n] },
\end{equation}
where $ I^{(t)}_{k}[n]  =P_n^{V2N\,(t)} { g}^{(t)}_{n \rightarrow k}[n] +\displaystyle\sum_{k^{\prime} \neq k} \eta^{(t)}_{k^{\prime}}[n]  P_{k^{\prime}}^{V2V\,(t)}[n]  g^{(t)}_{k^{\prime} \rightarrow k}[n]  $ is the collective interference at the  $k$-th V2V receiver from the $n$-th V2N link  and the $k^{\prime}$-th V2V links sharing the $n$-th sub-band. The code block length can be expressed by $m=B_w \Delta_{T}$, where $\Delta_{T}$ represents the packet transmission latency defined as the time taken for an entire packet of $L$ bits to be transmitted.   
In our work, packet transmission latency is fixed to
 1 milliseconds to ensure the latency requirement defined in 3GPP TR 36.885 \cite{3gpp}. Then, for a given payload size of $L$ bits per vehicle, the coding/data rate for the $k$-th V2V link,  $\Tilde{r}_{}=\frac{L}{B_w \Delta_{T}}$.
By rearranging Eq. (\ref{v2v_rate}), the decoding error probability at the $k$-th V2V receiver can be expressed as
\begin{equation}\label{eq:error}
\begin{aligned}
\varepsilon^{(t)}_{k}[n] \approx f_{Q}\left(\ln 2\sqrt{\frac{ B_w \Delta_{T} }{V^{(t)}_{k}[n]}} \left(C\left(\gamma_{k}^{V2V\,(t)}[n] \right)-\frac{L}{B_w \Delta_{T}}\right) \right).
\end{aligned}
\end{equation}
Then, based on the joint coding theory  \cite{wang2020packet},  the achievable  rate  of  the $k$-th V2V link can be expressed  as
\begin{equation} \label{rate}  \mathcal{T}^{(t)}_{k}=\displaystyle\sum_{n=1}^N \eta_k^{(t)}[n]\,\bigg( R_{k}^{V2V\,(t)}[n]\left(1- \varepsilon^{(t)}_{k}[n]  \right)  \bigg).
\end{equation}


\subsection{Problem Formulation}\label{sec:opt_prob} 
To maximize the sum-rate for the V2N  and V2V links while ensuring reliable information transmission for the V2V links, we jointly optimize the spectrum allocation $\eta^{(t)}_{k}[n]$ and the V2V transmission power $P_{k}^{V2V\,(t)}[n]$,  $\forall k \in \mathcal{K}$ and $\forall n \in \mathcal{N}$. Specifically, the sum-rate performance metric measures the total data rate achieved by all V2V and V2N communication links in the network. This metric is crucial in V2X communications because it directly reflects the network's ability to efficiently utilize the available spectrum while supporting multiple simultaneous links. A higher sum-rate indicates better spectral efficiency for data-intensive applications
and reliable
transmission of safety-critical messages.  Considering these  QoS requirements in vehicular networks, the objective function of our work is to jointly optimize the sum rate for the V2N links and the V2V links. The dynamic multi-objective resource allocation problem at time slot $t$ is formulated  as 
\begin{subequations}\label{opt_problem}
\begin{align}
\underset{\boldsymbol{\eta^{(t)}}, \boldsymbol{P^{(t)}}}{\operatorname{maximize}} & \left\{\mathrm{\omega_1} \displaystyle\sum_{n=1}^N R_n^{V2N\,(t)} + \mathrm{\omega_2} \displaystyle\sum_{k=1}^K \mathcal{T}^{(t)}_{k}\right\}  \label{obj}\\
\text{subject to} &  \nonumber \\
& 0 \leq \epsilon^{(t)}_k[n] \leq \varepsilon_{\max , k} \quad \forall k \in \mathcal{K}, \, \forall n \in \mathcal{N} \label{c1}  \\ 
& 0 \leq P_{k}^{V2V\,(t)}[n] \leq P_{\max }, \quad \forall k \in \mathcal{K}, \, \forall n \in \mathcal{N}  \label{c2}\\ 
& \sum_{n \in \mathcal{N}} \eta^{(t)}_{k}[n]=1, \quad \forall k \in \mathcal{K} \label{c3}\\ 
& \eta^{(t)}_{k}[n] \in\{0,1\} \label{c4}. 
\end{align}
\end{subequations}
where $\omega_1$ and $\omega_2$ represent the weights to balance the sum-rate of V2N and V2V link, respectively;  and 
$\boldsymbol{\eta}^{(t)}=$ $\Big[\eta_{1}[1], \ldots, \eta_{k}[n], \ldots, \eta_{K}[N]\Big] $, $\mathbf{P}^{(t)}=$ $\left[P_{1}^{V2V}[1], \ldots, P_{k}^{V2V}[n], \ldots, P_{K}^{V2V}[N]\right]$
are the vectors for the sub-band selection and the power allocations at time slot $t$, respectively. Then, in (\ref{c1}), $\epsilon_{\max, k}$ is the maximum error probability constraint for the $k$-th V2V link to ensure a predefined quality of service (QoS). (\ref{c2}) is the maximum power constraint for the $k$-th V2V transmitter. (\ref{c3}) indicates that each V2V link can only use one sub-band at a time.  

The optimization problem (\ref{opt_problem}) is a non-convex mixed integer nonlinear programming (MINLP) problem and involves sequential decision-making over the time slots. Due to the dynamic nature of the C-V2X networks and the associated short coherence time interval, centralized solutions are inadequate due to difficulties in acquiring the instantaneous CSI of all links \cite{myconf}. To address these issues, we propose to exploit DRL to solve the problem (\ref{opt_problem}) in a decentralized fashion. Specifically, the sequential decision-making over time can be encapsulated within a  Markov decision process (MDP) framework, and solved using a DRL-based strategy by treating each V2V transmitter as an agent interacting with
the unknown environment to maximize the reward. In what follows, we transform the problem of joint spectrum and power allocation into a multi-agent DRL problem.

\section {Multi-agent Reinforcement Learning-based Solution} \label{MARL-solution}
In this section, we design a DRL-based algorithm to solve the resource allocation problem in (\ref{opt_problem}). To build the foundations for the proposed DRL-based strategy, the description of key components of the RL framework in terms of the state space, action space, and reward function is first provided, followed by the proposed MADRL algorithm.

\subsection{Problem Transformation into DRL Framework } \label{drl_transformation}
DRL effectively addresses complex decision-making problems by training an agent to interact with its environment and learn optimal behaviors over time through continuous feedback and adjustment. The agent refines its actions to achieve optimal outcomes within a MDP framework \cite{DRL_basics}. An MDP is encapsulated by a tuple  \((\mathcal{S}, \mathcal{A}, \mathcal{P}, \mathcal{R})\), where \(\mathcal{S}\) is a set of states, \(\mathcal{A}\) is a set of actions, \(\mathcal{P}\) is the transition probability from states to actions, and \(\mathcal{R}\) is the reward for taking action in a state. At each discrete time step \(t\), the agent observes the state \(s^{(t)}\), takes action \(a^{(t)}\) according to policy \(\pi(s^{(t)}, a^{(t)})\), transitions to state \(s^{(t+1)}\), and receives a reward \(r^{(t)}\). The DRL-based system design aims to learn an optimal policy that maximizes the expected return \cite{R12}, \cite{Nasir_TVT}. Therefore, we cast the optimization problem  (\ref{opt_problem}) as an MDP by mapping the key elements from the DRL framework to the resource allocation problem.

In the multi-agent scenario, each V2V transmitter acts as an agent and interacts with the unknown communication environment. The joint action $\bar{\mathcal{A}}^{(t)}=\left(a_1^{(t)},\cdots, a_k^{(t)}, \cdots a_K^{(t)}\right)$, with $a_k^{(t)}$ as the $k$-th V2V agent's action in the time slot $t$, collectively influences the common environment and directs the agents towards the optimal policy. Hence, in our MADRL framework, different agents compete for limited power and spectrum resources, and the resource-sharing problem is shifted to a fully cooperative one by providing all agents with the same common reward. Next, we introduce the key components of the DRL framework in detail in terms of the state space, action space, and reward function.

\subsubsection{\textbf{State Space}}  The state of the $k$-th V2V agent at time slot $t$, ${s_k}^{(t)}$, includes the channel information and the received interference from other links. Specifically, the channel information for the $k$-th V2V agent includes the instantaneous channel gain of its own link  ${g}^{(t)}_{k}[n]$, $\forall n \in \mathcal{N}$, the interference channel gain
from the transmitter of the $n$-th V2N link and the $k^{\prime}$-th V2V link, ${ g}^{(t)}_{n \rightarrow k}[n]$ and  $g^{(t)}_{k^{\prime} \rightarrow k}[n]$,\, $\forall n \in \mathcal{N}$,\,  ($k^{\prime} \neq k$), and the interference channel gain from its transmitter to the BS ${ g}^{(t)}_{k \rightarrow B}[n]$, $\forall n \in \mathcal{N}$. 
The channel information $g^{(t)}_{k}[n]$, ${ g}_{n\rightarrow k}^{(t)}[n] $, and $g_{k^{\prime}\rightarrow k}^{(t)}[n]$, can be accurately estimated by the receiver of the $k$-th V2V link at the beginning of each scheduling slot, and we assume it is also available instantaneously at the transmitter through delay-free feedback. 
The CSI for $g_{k\rightarrow B}^{(t)}[n]$ is estimated at the BS and broadcast to all vehicles within the BS coverage at the beginning of each scheduling slot \cite{DRL23_7}, \cite{R12}. 

Additionally, the aggregate interference from the V2N links and the V2V links over the $n$-th  sub-band in the past time slot ${I}_{k}^{(t-1)}[n]$ is included in the state space of the $k$-th agent, which is expressed as 
\begin{equation}
 \nonumber {I}_{k}^{(t-1)}[n]=P_n^{V2N{(t-1)}} { g}^{(t)}_{n \rightarrow k}[n]+\displaystyle\sum_{k^{\prime} \neq k} \eta_{k^{\prime}}[n] P_{k^{\prime}}^{V2V {(t-1)}}[n] g_{k^{\prime}\rightarrow k}^{(t)}[n],
\end{equation}
where the first term is the interference from the  V2N transmitter and the second term is the aggregate interference from all V2V transmitters using the same $n$-th sub-band in the previous time slot. It should be noted that the sum interference for the $k$-th V2V agent is estimated in slot $t$ as follows. At the beginning of the current time slot, the new transmit power and spectrum allocation decision of each V2V agent has not been determined and each V2V agent utilizes the resource decision of the previous time slot to calculate ${I}_{k}^{(t-1)}[n]$, although the CSI of the whole network has changed. Then, the current state $s_k^{(t)}$  of the $k$-th V2V agent is given by

\begin{equation}
        {s}_k^{(t)}= \Biggl\{\left({g}^{(t)}_{k}[n], g^{(t)}_{k^{\prime} \rightarrow k}[n], {g}^{(t)}_{k \rightarrow B}[n], { g}^{(t)}_{n \rightarrow k}[n],\,
        {I}_{k}^{(t-1)}[n]\right) \mid {n \in N}\Biggr\}.
\end{equation}

The scale of the state space is $(K+2) \times N$ per V2V  agent, {which includes $N$ elements for the direct link between the $k$-th V2V transmitter and its receiver, $(K \times N -N)$ elements for the interfering channels from other V2V transmitters, $N$ elements for the V2N links, and $N$ elements for the sum interference}.

\subsubsection{\textbf{Action Space}} The action of each agent at time slot $t$ corresponds to  the spectrum sub-band  and transmission power selection. While $N$ disjoint sub-bands are preoccupied by the $N$ V2N links and all V2V links share these sub-bands, each V2V agent can only pick at most one spectrum sub-band during the same time slot $\left(\sum_{n \in \mathcal{N}} \eta^{(t)}_{k}[n]=1, \forall \mathcal{K}\right)$. The power allocation is discretized in $L_Q$ levels given by $\mathcal{P}=\left\{\frac{P_{\max }}{L_Q}, \frac{2 P_{\max }}{L_Q}, \cdots, P_{\max }\right\}$, with  $\mathcal{P}$  representing the power action set. 
Then, the  action executed by the $k$-th V2V agent in the time slot $t$ can be expressed as  $a_k^{(t)}=\left[\eta^{(t)}_{k}[n],\,P_{k}^{V2V\,(t)}[n]\right]$, which includes $\eta^{(t)}_{k}[n]$ and $P_{k}^{V2V\,(t)}[n]$ for spectrum sub-band and power action decision, respectively. Accordingly, each agent has an action vector with the dimension of $L_Q \times N$.

\subsubsection{\textbf{Reward Function}}
A common reward is designed for all agents to encourage cooperation among the multiple V2V agents. In order to reflect the performance of the decision taken by the agents, the immediate reward function is formulated as
\begin{equation}
\begin{aligned} \label{reward}
r^{(t)}=&\mathrm{\lambda_1} \displaystyle\sum_{n=1}^N R_n^{(t)\,V2N} + \mathrm{\lambda_2} \displaystyle\sum_{k=1}^K \mathcal{T}^{(t)}_{k}\\
&-\lambda_3 \displaystyle\sum_{k=1}^K \left(\max \left(\epsilon_{k}^{(t)}-\epsilon_{max,k}, 0  \right)\right),
\end{aligned}
\end{equation}
where the summation terms respectively represent the sum-rate of V2N links, the sum-rate of V2V links,
and the reliability requirements of the V2V links; and $\lambda_{i}, \: i \in\{1,2,3\}$, are the positive weights to balance the utility and the penalty cost in terms of constraint violations.  To align the designed reward function at each step with the network objective (\ref{obj}), the first two terms in (\ref{reward}) are included as positive reward elements. Additionally, the reward function incorporates a penalty for failing to satisfy the reliability constraint related to the decoding error probability. The selection of weight coefficients is critical, as they significantly affect the learning efficiency and convergence of the algorithm. In practical scenarios, these coefficients, denoted as $\lambda_{i}$, are considered hyperparameters and require empirical tuning. Setting equal positive weights for all reward components prioritizes maximizing the sum rate of V2V and V2N links. However, our tuning experience suggests such a design will impede the algorithm’s convergence as the agents struggle to learn reliability aspects. This is because the penalty term, based on target error probabilities, is much smaller than other reward components. The reward function components are normalized by setting $\lambda_{1}  = \frac{1}{5K}$, $\lambda_{2} = \frac{1}{R^{'}}$, where $R^{'}=\sum_{k=1}^K \sum_{n=1}^N C\left(\gamma_{k}^{V2V\,(t)}[n]\right)$ is the sum V2V Shannon rate and $\lambda_3 = 1$. This hyperparameter selection approach balances the contribution of utility and constraint violation cost while enabling stable updates of the  Q-values for improved convergence.

\subsection{Proposed Multi-Agent DRL-based Algorithm}\label{MADRL_proposed}

\begin{figure}[t] 
 \centering
\includegraphics[width= 0.9\linewidth]{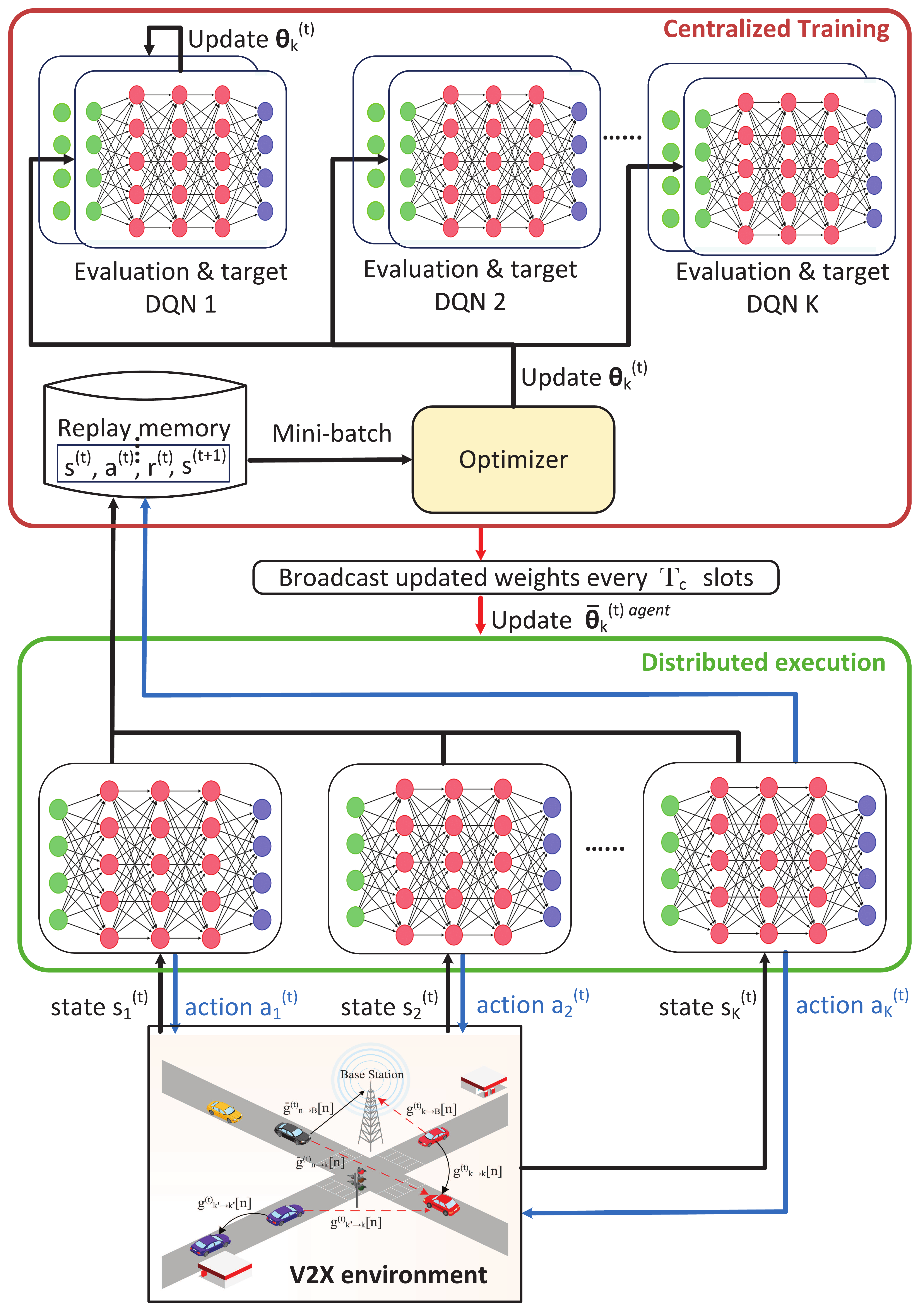}
\caption{ Illustration of the proposed multi-agent deep reinforcement learning
algorithm.} 
\label{fig_DRL}
\end{figure}

We propose an MADRL scheme with each V2V transmitter as an agent. The multi-agent scenario breaks the non-stationarity assumption of the MDP as the environment may not be stationary from the agent's perspective as other learning agents update their policies. To effectively handle the non-stationarity issue in multi-agent settings, we propose the centralized training and decentralized implementation approach.  Specifically, DQNs are distributively executed at the V2V transmitters, whereas the weight parameters are centrally trained at the BS  to ease implementation and improve stability. 

The proposed centralized-trained distributively-executed framework is shown in Fig. (\ref{fig_DRL}). The  centralized training  and distributed execution procedures are detailed as follows: 

\subsubsection{ \textbf{Centralized Training}}
In the training phase, $K$ evaluation DQNs $Q\left(s_{k}^{(t)},a_{k}^{(t)};\theta_{k}^{(t)}\right)$ with weight parameters $\theta_k^{(t)}$, and $K$ corresponding target DQNs $Q\left(s_{k}^{(t)}, a_{k}^{(t)};\bar{\theta}_k^{(t)}\right)$ with weight parameters $\bar{\theta}_k^{(t)}$ are established at the BS, whereas a local DQN  $Q^{\,agent}\left(s_k^{(t)},a_k^{(t)}; \bar{\theta_k}^{(t)\, agent}\right)$, with weight parameters $\bar{\theta_k}^{(t)\, agent}$ is established at the vehicle transmitter as the $k$-th V2V agent. The local network $Q^{\,agent}\left(s_k^{(t)},a_k^{(t)}; \bar{\theta_k}^{(t)\, agent}\right)$  has the same structure as the corresponding evaluation network $Q\left(s_{k}^{(t)},a_{k}^{(t)};\theta_{k}^{(t)}\right)$, but is established at the vehicle transmitter instead. The input
and the output of each V2V agent are the local state
information and the adopted local action (i.e., transmit
power and spectrum allocation), respectively. In  time slot $t$,  each  V2V agent $k$  observes 
its current state $s_k^{(t)}$ and independently executes the  action ${a}_k^{(t)}$ following the  $\epsilon$-greedy policy. The local observations $\left(s_{k}^{(t)}, a_{k}^{(t)}\right)$ of each V2V agent are uploaded to the BS at the beginning of time slot $t$, whereas the common reward $r^{(t)}$ is evaluated at the BS.  Afterwards,  the  experience  $e_{k}^{(t)}=\left\{s_{k}^{(t)}, a_{k}^{(t)},r_{k}^{(t)},s_{k}^{(t+1)}\right\}$ of each V2V agent $k$ is stored in the replay buffer memory $\mathcal{D}$. Note that all
V2V agents share the same reward in the system such that cooperative behavior among them is encouraged, and the selected experiences may include the experiences of different agents at different time slots since a single replay buffer memory is used to store experiences for all agents.  In this manner, the BS trains the evaluation DQNs using the experiences gathered from all agents. Then, by sampling a mini-batch of $D$ experiences from $\mathcal{D}$ in every $T_{b}$ time slot, the trainer optimizes the weights for the DQNs by minimizing the loss function using the gradient descent technique:
 \begin{equation}\label{loss_DQN}
 \begin{aligned}
     \mathcal{L} 
     &= \frac{1}{\|D\|}\sum_{e\,\in\,{D}}\left(y^{(t)}-Q(s^{(t)}, a^{(t)} ; {\theta^{(t)}})\right)^{2},
 \end{aligned}
 \end{equation}
where  $y^{(t)}=r^{(t+1)}+\zeta\, \underset{a^{(t+1)}}{\max} Q\left(s^{(t+1)},a^{(t+1)};  \bar{{\theta}}^{(t)}\right)$ is the target $Q$-value generated by the target network. In every $T_{c}$ time slots, the BS updates the target DQNs with the weights of the evaluation DQNs and broadcasts the latest trained weight parameters ${{\theta_k}^{(t)}}$ of the evaluation DQN to local agent to update its weights $\bar{{\theta_k}}^{(t)\, agent}, \forall k \in K$.  The complete training procedure is summarized in Algorithm \ref{DQN_algo_train}, which takes the initial DQN model for each V2V agent, batch sampling period $\mathcal{T}_{b}$, weight copy period $\mathcal{T}_{c}$, and the number of train episodes $\mathcal{E}_{train}$ as the input and outputs the real-valued trained network's weight vector $\boldsymbol{\theta} = \big\{ {\theta_k} \big\}_{k=1}^K$ along with the background dataset $\mathcal{X}_{BG}$ to be used in Section \ref{XAI} for explainability purpose. Algorithm \ref{DQN_algo_train} achieves convergence upon completion of the training process.

 \begin{algorithm} 
  \SetAlgoLined 
  \caption{Centralized Training Stage }\footnotesize
  \label{DQN_algo_train}
  \KwIn{Initial DQN models, $\mathcal{T}_{b}$, $\mathcal{T}_{c}$,  $\mathcal{E}_{train}$}
  \KwOut{Trained weights $\boldsymbol{\theta}$, background dataset $\mathcal{X}_{BG}$}
  Initialize  replay memory $\mathcal{D}  \leftarrow \emptyset$,  and $\mathcal{X}_{BG}  \leftarrow \emptyset$\;
  Random initialize the weights $\theta_k^{(t)}$ of DQNs, and the weights of target  DQNs  as $\bar{\theta}_k^{(t)}=\theta_k^{(t)},  \forall k \in \mathcal{K} $\;
  \For{ $e=0:\mathcal{E}_{train}$ }{
    Generate the V2N and V2V communications links\;
    \For{$t=0:\mathcal{T}_{steps}$ }{
        \For{$k=0:K$ }{
          Observe the state ${s}_k^{(t)}$\;
          Choose the joint action $a_k^{(t)}$ following the $\epsilon$-greedy policy based on ${s}_k^{(t)}$\;
          Evaluate the reward $r^{(t+1)}$ in (\ref{reward}) and move to next state ${s}_k^{(t+1)}$ by executing  action $a_k^{(t)}$\;
          $\mathcal{D}  \leftarrow \mathcal{D} \cup \left\{{s}_k^{(t)}, a_k^{(t)}, r_k^{(t)}, {s}_{k}^{(t+1)}\right\}$\;
          $\mathcal{X}_{BG}  \leftarrow \mathcal{X}_{BG} \cup \left\{{s}_k^{(t)},  a_k^{(t)} \right\}$\;
          \If{$t \, > \,T_{b}$}{
            Randomly sample a mini-batch of experiences from $\mathcal{D}$\;
            Update weights ${\theta_k}^{(t)}$ by minimizing the corresponding loss function in (\ref{loss_DQN})  via the gradient descent technique\;
            \If{$t \, \% \, T_{c}=0$}{
                Copy  weights ${{\theta_k}^{(t)}}$ of training DQN  to
                 target DQN $\bar{{\theta_k}}^{(t)}$\; 
                 Update weights $\bar{{\theta_k}}^{(t)\,agent}$ of local agent with latest weights ${{\theta_k}^{(t)}}$ of training DQN\;
            }
           }
         }
     }
  }
 \KwRet{$\boldsymbol{\theta}$, $\mathcal{X}_{BG}$}
\end{algorithm} 

\subsubsection{\textbf{Distributed Execution}}
 During the distributed implementation phase,  the $\epsilon$-greedy algorithm is terminated, i.e., agents stop exploring the environment. Each V2V agent observes its environment state $s_k^{(t)}$ and executes action ${a}_k^{(t)}$ determined by the maximum action value according to its trained DQN, i.e., $a_k^{(t)}={\arg\max_{a^{(t)}_k} }\,Q^{\,agent}\left(s_k^{(t)},a_k^{(t)}; \bar{\theta_k}^{(t)\, agent}\right)$. Consequently, all V2V links start transmission with the power level and spectrum sub-band determined by their selected actions. Further, the distributed execution phase outputs a hold-out dataset $\mathcal{X}$ containing the environment states and a copy of the action values ($Q$-values) as the labels. Note that the predicted action of a discrete action space is the index of
the $\textit{argmax}$ function, and the experience replay already contains a copy of the action predicted by
the trained network. We relegate the detailed design and use of datasets $\mathcal{X}$ and $\mathcal{X}_{BG}$ to Section \ref{XAI}.

\section {XAI Guided Feature Selection  and \\ Model Simplification}\label{XAI}
In this section, we introduce our XAI-based approach to simplify the trained DRL models' complexity by quantifying the relevance of input state features to the output actions for our proposed DRL-based V2X communication system. A novel post-hoc and model-agnostic framework is proposed to explain and reduce the complexity of the trained networks by eliminating some of the input variables using the state feature importance ranking.

As the basis of our methodology, we leverage the  SHAP \cite{Deepshap}, a unified approach that offers a strong theoretically grounded framework for computing local and global explanations of a machine learning model. In what follows, we first introduce Shapley values, followed by a brief description of SHAP. Then,  our proposed XAI-based systematic methodology and input state feature selection algorithm for model simplification are presented.

\subsection{ Preliminaries on  Shapley Values and SHAP}
Feature importance rankings necessitate the computation of  Shapley values which can be interpreted as the average expected marginal contribution of each input feature \( x_l \in \mathbf{x} \) to a model's prediction \( f(\mathbf{x}) \), where $\mathbf{x}$ is the vector of original feature values. The Shapley value for \( x_l \) is computed by considering all possible coalitions of features and calculating the weighted average difference in predictions with and without the feature \( x_l \). For a prediction model  \( f(\cdot) \) with input $\mathbf{x}$, the Shapley value for feature \( x_l \) is given by:
\begin{equation}\label{shapley_equation}
\phi_l(f, \mathbf{x}) = \sum_{\mathbf{z'} \subseteq \mathbf{x'} } \frac{|\mathbf{z'} |!(L - |\mathbf{z'}| - 1)!}{L!} \left[ f(\mathbf{z'}) - f(\mathbf{z'}  \setminus \{x_l\}) \right]
\end{equation}
where \( L \) is the total number of original features, $\mathbf{z'}$ is the subset of features used in the model,  \( f(\mathbf{z'}) \) is the model's prediction on feature subset \(\mathbf{z'} \), $f(\mathbf{z'}  \setminus \{x_l\})$ is the model's prediction on feature subset \( \mathbf{z'}  \) excluding feature $x_l$. When calculating  \( f(\mathbf{z'}) \), the  $x_l$-th feature is masked out and then
simulated by drawing random instances of the  $x_l$-th feature from the background dataset. The Shapley value $\phi_l(f, \mathbf{x})$  signifies the impact of feature $x_l$ on transitioning from the reference output value, i.e., the expected value without knowledge of the $x_l$-th feature values to the actual output provided by the prediction model. These contributions carry both magnitude and sign, allowing to assess a feature's significance. However, exactly calculating Shapley values requires searching through all possible $2^L$ feature combinations while retraining models, which is computationally prohibitive. As a remedy, SHAP provides a computationally efficient way to calculate Shapley values by iterating over a small subset of possible feature permutations using sampling approximations \cite{Deepshap}. It combines previously proposed explainability methods, such as LIME \cite{lime} and Deep Learning Important FeaTures (DeepLIFT) \cite{lift}, to approximate the Shapley values. As an additive feature attribution method, SHAP approximates the output $f(\mathbf{x})$ of the original complex model by  summing the scores attributed to each feature, $\phi_l$, as follows:
\begin{equation}\label{shap_equation}
g\left(\mathbf{z}^{\prime}\right)=\phi_0+\sum_{l=1}^L \phi_l z_l^{\prime}
\end{equation}
where $g\left(\mathbf{z}^{\prime}\right) \approx f(\mathbf{x}) $ is the explanation model, and \( \phi_0 \)  corresponds to the average or expected output of the model seen during training that
can be interpreted as the initial output of the model before the impact of any features is considered. Moreover, Deep Shapley additive explanation (Deep-SHAP) is an extended version designed specifically for DNNs and adapts DeepLIFT algorithm to approximate Shapley values by linearizing the non-linear components of DNNs \cite{Deepshap}.  Deep-SHAP adds explainability to the DNNs by decomposing the output prediction of  DNNs on a specific input by backpropagating the contributions of all neurons to every input feature.  Instead of using a single reference value, Deep-SHAP exploits a distribution of background samples such that the resulting Shapley values sum up to the difference between the expected model output on the passed background samples and the current model output. 

\subsection{Proposed Systematic XAI Methodology}\label{XAI_method}
The systematic XAI methodology offers two key benefits: First, it simplifies the input size by eliminating non-influential features through an iterative SHAP-based feature selection algorithm. Second, it reduces the DNN model's complexity by adjusting the architecture based on the refined input size, thereby lowering computational load and minimizing parameter updates for the distributed implementation. Our methodology can be divided into two stages. The first
stage involves generating state feature importance ranking of
the trained models using the Deep-SHAP explainer. The second stage uses importance-based rankings to simplify the state space of the agents using a novel variable feature selection method.  

\begin{figure*}[t]
    \centering
    \includegraphics[width=0.9\linewidth]{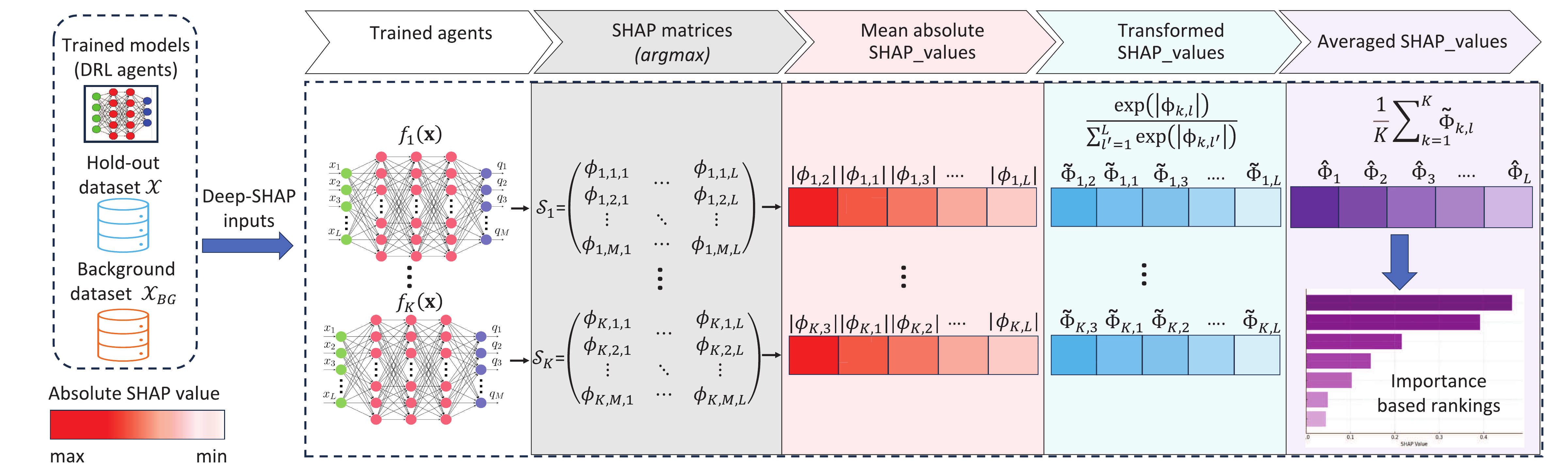} 
    \caption{
The process of generating SHAP\_values for multiple trained models using Deep-SHAP. The pre-trained models $f_k(\mathbf{x}), \forall k \in K$ and the background dataset $\mathcal{X}_{BG}$ are utilized to calculate SHAP\_values for each instance in the hold-out dataset $\mathcal{X}$. The $\textit{argmax}$ output helps select the appropriate vector of SHAP\_values. Subsequently, the mean-absolute SHAP\_values are calculated and  sorted in descending order to encode feature importance. These values are then transformed using the softmax transformation. The transformed SHAP\_values are averaged across the $K$ trained models. Here, $L$ is the total number of state features, and $M$ is the number of valid actions.
}
    \label{SHAPflow}
\end{figure*}

For notation consistency, we denote the $k$-th trained V2V agent as the prediction model, i.e.,  $ f_k(\mathbf{x}) \equiv   Q_k^{\,agent}\left({\mathbf{x}}\right)$
, which takes real-valued state feature vector $\mathbf{x}=\left[x_1, \ldots, x_L\right]$ as an input, where $L=(K+2) \times N$ is the number of input neurons. 
The prediction model $ f_k(\mathbf{x})$  outputs the $Q$-value of each
possible action denoted by $\mathbf{q}_k=\left[q_{k,1},\ldots,  q_{k,m},  \ldots, q_{k,M}\right]$, where $M=L_Q \times N$ 
is the number of output layer neurons and equal to the cardinality of the action set described in Section \ref{drl_transformation}. 
The trained model uses  $\textit{argmax}$ operator to select the action maximizing the model's output.

In the first stage of generating state feature importance ranking using the trained agents, we instantiate a DeepExplainer from the SHAP library \cite{lundberg} and pass the trained agent’s DQNs along with the background dataset to it. The background dataset $\mathcal{X}_{BG}$ is generated during the training process and serves as a prior expectation for the sample instances to be explained \cite{R19}. To compute the SHAP\_values, a hold-out dataset $\mathcal{X}$ is used, which contains the state features and a copy of all possible actions, i.e., $Q$-values. This dataset is generated by evaluating the performance of the well-trained agents over a set of test episodes.
Deep-SHAP produces local explanations in the form of importance score or "SHAP\_values," $\boldsymbol{\phi}_k(\mathbf{x})= \left[\phi_{k,1}, \ldots, \phi_{k,L}\right]$, where  $\phi_{k,l}$ is the SHAP\_value associated with feature $x_l \in \mathbf{x}$  and probable action  $q_{k,m}$ of the $k$-th prediction model. To obtain global explanations, SHAP\_values are obtained by repeating this across the entire database,  resulting in a SHAP matrix for each sample instance in $\mathcal{X}$. The process for calculating the SHAP\_values for the prediction model $f_k(\mathbf{x})$  using Deep-SHAP is delineated in Fig. \ref{SHAPflow}, wherein Deep-SHAP first computes the SHAP matrix $\mathcal{S}_{k}$ for state sample $\mathbf{x} \in \mathcal{X}$  with one row for each action value $q_{k,m}$ and one column per state feature $x_l\in \mathbf{x}$.  Then, the SHAP\_values corresponding to the maximizing action index are selected from  $\mathcal{S}_{k}$.
Afterward,  the absolute mean of the feature column of $\mathcal{S}_{k}$ is calculated across all data instances in $\mathcal{X}$. The resulting vector of mean absolute SHAP\_values $\left|\boldsymbol{\phi}_{k}(\boldsymbol{\mathbf{x}})\right|= \left[\left|\phi_{k,1}\right|, \ldots, \left|\phi_{k,L}\right|\right]$ is sorted in descending order. The first position of the resulting vector contains the most important feature, the second position contains the second most important, and so on.  To reveal the meaning behind the SHAP\_values in terms of their contribution to the models' outputs,  we transform  the SHAP\_values into a probabilistic distribution $\mathbf{\Tilde{\Phi}}_{k} \in \left[ 0, 1\right]^L$ through the following softmax transformation:
\begin{equation} \label{softmax_transform}
    \mathbf{\Tilde{\Phi}}_{k,l}=\frac{\exp \left\{\left|\phi_{k,l}\right|\right\}}{\sum_{l^{\prime}=1}^L \exp \left\{\left|\phi_{k,l^{\prime}}\right|\right\}},   \forall k \in \mathcal{K}, l^{}=1, \ldots, L,
    \end{equation}
where $\mathbf{\Tilde{\Phi}}_{k,l} = \left[\Tilde{\Phi}_{k,1}, \ldots, \Tilde{\Phi}_{k,L}\right]$ denotes the real-valued vector of transformed SHAP\_values  for the $k$-th  prediction model.

To obtain global state feature importance ranking,  the transformed absolute SHAP\_values are obtained for each trained agent using the above procedure and averaged across the $K$ trained models. The averaged transformed SHAP\_values, denoted by $\mathbf{\widehat{{\Phi}}}= \left[\widehat{{\Phi}}_{1}, \ldots, \widehat{{\Phi}}_{L}\right]$  are utilized to rank the input state features, which can be interpreted as global state-feature importance scores. Then, by utilizing an appropriate feature selection strategy, the most important state elements contributing to the model outcome are selected, and the model is retrained using the subset of these important states. Note that the presented systematic feature selection strategy in  Fig. \ref{SHAPflow} is model-agnostic and applicable to any other type of pre-trained  DRL-based radio resource management system.

\subsection{Post-hoc  SHAP-based Input State Selection Algorithm }
We devise an iterative state feature selection algorithm based on SHAP-based state feature importance rankings.  The post-hoc  SHAP-based state feature selection strategy is summarized in Algorithm \ref{algo2}, whose inputs are the $K$ trained agents, hold-out dataset $\mathcal{X}$, background dataset $\mathcal{X}_{\text{BG}}$, and the precision threshold $\Delta$, whereas the outputs are the optimal subset of features retained using the proposed feature selection strategy.  The developed variable feature selection strategy selects the number of important state features
by repeatedly evaluating the trained models based on a predefined metric while masking step-by-step the values of the $p$ least important features according to the given importance-based feature ranking. Significant changes in the evaluation results indicate that a sufficiently important state feature has been randomized in the current step. Consequently, all state features randomized in the previous step can be confidently removed from the trained network's input.

While the network is trained to maximize cumulative reward in (\ref{reward}), this reward is not the most illustrative way to evaluate its post-hoc performance due to its high variance.
To evaluate the trained models' performance, we consider the following reward value averaged across the hold-out dataset \begin{equation} \label{eq:ANR}
\alpha = \frac{1}{\|\mathcal{X}\|} \sum_{i = 1}^{\|\mathcal{X}\|} r^{(i)}, 
\end{equation}
where $r^{(i)}$ is calculated using the $i$-th state-action tuple in the hold-out data set, $\mathcal{X}$, as per (\ref{reward}).  

\begin{algorithm}[t!]
\SetAlgoLined
\caption{Iterative SHAP-based Feature Selection}\footnotesize
\label{algo2}
  \KwIn{$f_{k \,(\forall k \in K)}$, $\mathcal{X}$, $\mathcal{X}_{BG}$, $\Delta$}
\KwOut{Subset of input state features}
\For{$k=0,1,\ldots,K$}{
    Generate SHAP matrix $\mathcal{S}_{k}$ for $f_k(\mathbf{x})$ using $\mathcal{X}$ and $\mathcal{X}_{BG}$\;
    Obtain transformed SHAP\_values $\mathbf{\Tilde{\Phi}}_{k,l}$ using (\ref{softmax_transform})\;
}
$\mathbf{\widehat{\Phi}} \gets $ Average $\mathbf{\Tilde{\Phi}}_{k,l}$ across the $K$ agents\;
Generate  global state-feature importance rankings using $\mathbf{\widehat{\Phi}}$ \;
$\alpha_{{original}}\gets $ Evaluate the average network performance on $\mathcal{X}$ using (\ref{eq:ANR})\;
$p \leftarrow 1$\;
\While{$p < L$}{ 
    Mask the values of the $p$ least important features within $\mathcal{X}$\; \label{stepback}
    $\alpha_{{simplified}} \gets $Re-evaluate the average network performance using the partially masked dataset \;
    \If{$|\alpha_{{original}} - \alpha_{{simplified}}| \geq \Delta$}{
        Eliminate the $p-1$ least important features\;
        \KwRet{$\textit{Subset of input state features}$}\; }
    \Else{
     $p \leftarrow p+1$\;
    }
}
Eliminate the $p-1$ least important features\;
\KwRet{Subset of input state features}
\end{algorithm}
 
Algorithm \ref{algo2} starts with a for loop (Lines $1-4$) that iteratively
computes the SHAP matrix $\mathcal{S}_{k}, \: \forall k\in K,$ and obtains transformed SHAP\_values, $ \mathbf{\Tilde{\Phi}}_{k,l},\: \forall k \in K,$ as per (\ref{softmax_transform}). 
Then, $\mathbf{\Tilde{\Phi}}_{k,l}$,  $\forall k \in K,$  are averaged across the trained models and utilized to rank the input state features (Lines $5-6$). The generated feature rankings are used to select a subset of state features using the average network performance metric in (\ref{eq:ANR}). The algorithm first evaluates average network performance in (\ref{eq:ANR}) with no state features removed, denoted by $\alpha_{original}$  (Line $7$).    Then, the $p$ least important features are masked according to SHAP importance-based feature ranking, i.e., their values are changed to the respective column variance (Line $10)$. The trained models are re-evaluated on the modified dataset, and the evaluation results are denoted as $\alpha_{simplified}$ (Line $11$). If there is a significant change in the evaluation results, i.e., $\left|\alpha_{{original}}-\alpha_{{simplified}}\right| \geq \Delta$, where $\Delta$ is the precision threshold, then the  $p-1$ least important state features are removed and only $L-p+1$  influential state features are retained (Lines $12-15$). Otherwise, the value of $p$ is incremented, and the state feature masking and re-evaluation steps previously defined are performed iteratively (Lines $17-19)$. It should be noted that the value of $\Delta$ is set to be between 1-10\% of the average network performance metric, ensuring that the algorithm outputs only a subset of importance-based features.

 One major obstacle associated with the Deep-SHAP based  SHAP\_values estimation is the considerable computational complexity, which increases exponentially with the number of state features and linearly with the number of background data samples \cite{Deepshap}. Our initial experimentation shows that utilizing the whole training dataset $\mathcal{X}_{\text{BG}}$ significantly slows down the SHAP computations.  To strike a balance between achieving high accuracy in SHAP\_values  estimates and keeping computation time reasonable, we utilize the transformation in the existing SHAP implementation by $\mathit{K}$-means to summarize $\mathcal{X}_{BG}$ \cite{lundberg}. Specifically, we use 1\%  $\mathit{K}$-means summary, i.e., 4,000 samples of  $\mathcal{X}_{BG}$ and  10\%  $\mathit{K}$-means summary, i.e., 1,000 test samples,  to serve as the background and hold-out dataset, respectively. This summarization is sufficient to accurately assign state feature importance to different input states without incurring the full computational burden of calculating SHAP\_values over all data samples in the background dataset.

Note that the fast-changing nature of the state features does not pose a problem for the proposed XAI method. The post-training inference process can be performed in an environment twin (a simulation of the real environment), and the feature selection process
is fast enough so that the environment does not change in the meantime. Additionally, our XAI-based feature selection process does not interact with the environment as it only utilizes the already-trained DRL agents and stored environment observations.

\section{Performance Evaluation}\label{sec:simulation}
This section evaluates the performance of the proposed MADRL resource allocation scheme in a single-cell V2X communications system. Numerical results are presented to validate the effectiveness of applying the XAI-based model simplification strategy.

The following algorithms are simulated for performance comparison to verify the efficiency of our proposed  XAI-based simplified multi-agent DRL (Simplified-MADRL) scheme.

\begin{enumerate}
    \item \textbf{Original multi-agent DRL (Original-MADRL)}: This algorithm trains the multiple V2V agents using the proposed centralized training with a decentralized execution framework described in Section \ref{MADRL_proposed}. This algorithm considers the original full state features of the multiple agents during the training and execution process, hence named Original-MADRL.

     \item \textbf{Centralized Single-agent DRL (SADRL)}: This algorithm trains a single meta-agent at the BS, which outputs joint actions for all V2V agents. Since a single DQN  jointly handles spectrum sub-band selection and transmit power control, the complexity of the output layer size and state action pairs to be visited for convergence is proportional to the total number of V2V pairs in the network. Apart from the complexity, this algorithm is executed in a centralized way with accurate global CSI availability requirements.

    \item \textbf{Full power}: This algorithm selects the channel with the lowest interference for V2V transmission, setting the transmit power to the maximum level \( P_{\text{max}} \) for all V2V links. The computational complexity is dominated by the enumeration of all  \( N \) available spectrum sub-bands, where a fast-sorting algorithm is used centrally at the BS to efficiently compare interference levels across the sub-bands and identify the one with the least interference \cite{DRL23_3}.

    \item \textbf{Random allocation}: This algorithm chooses a random transmit power and spectrum sub-band action at the beginning of each time step.

\end{enumerate}
 
For the learning algorithms, we consider an episodic setting with each episode lasting for $ \mathcal{T}_{steps}=$100 time steps. 
At the beginning of each episode, the environment state is initialized randomly, determined by the initial transmission powers of vehicular links and environment states. 
The training phase lasts $ \mathcal{E}_{train}=4000$ episodes, whereas $\mathcal{E}_{test}=1000$ episodes are dedicated to the testing phase, where the well-trained DNNs are utilized to evaluate the performance of the different learning algorithms.

All simulations are performed on a $10$-Core Intel(R) Xenon(R) Silver $4114$, $2.2 \mathrm{GHz}$ system equipped with an Nvidia Quadro P2000 graphics processing unit (GPU). The V2X simulation environment and algorithms are developed in Python, and the DNNs are built and trained using Tensorflow API. The open-source implementation of SHAP \cite{lundberg} is used to compute explanations for the models.

\subsection{Simulation Setup} \label{sec:Sim_setup}
We follow the simulation setup for V2X communications in annex A 3GPP TR 36.885, which describes in detail the vehicular channel models, vehicle mobility, and vehicular data traffic for an urban scenario \cite{3gpp}. The urban scenario considers an area of size 750x1299 meters with 9 buildings and roads between buildings, with 2 lanes in each direction for each road. Vehicles are moving in this area at equal constant speed. $N$  V2N links are initiated by $N$ vehicles and the $K$ V2V links are formed between each vehicle with its surrounding
neighbors. The V2N links follow the  path-loss model $PL(d)=128.1 + 37.6 \log _{10}(d)$,  where $d$ is the distance in $[\mathrm{km}]$ of the vehicle from the BS \cite{weiner}. For the V2V links, we use the WINNER path-loss model $PL(d)=A \log _{10}(d)+B+C \log _{10}\bigg(\frac{f_{c}}{5.0}\bigg)$, where $d$ is the  inter-vehicular distance in $[\mathrm{m}]$, $f_{c}$ is the system frequency in $[\mathrm{GHz}]$, the parameter $A$ describes the impact of path-loss exponent, the parameter $B$ is the intercept, and parameter $C$ controls the frequency dependency. We
set  $\omega_1 = \omega_2 =1$ in (\ref{opt_problem}) to consider a fair and equal contribution of  V2N and V2V link's sum rates. The parameter values for $A$, $B$, $C$ are calculated according to the guidelines (Table 4-4)  provided for $\textit{B}$1 urban micro-cell scenario \cite{weiner}. 
To achieve the demand of the URLLC, we set the  packet transmission latency $\Delta_{T}$ to $1$ ms, and the bandwidth occupied by each V2N link $B_w$ to $180$ KHz. Other important simulation parameters are provided in Table \ref{table_sim}.

The hyperparameter values for the DQNs are tabulated in Table \ref{table_DQN}. The tabulated parameters are fine-tuned to reach the desired results and chosen as the final settings because they consistently deliver the best performance across multiple experimental trials.
The initial learning rate $\delta^{(0)}=0.01$ is annealed as $\delta^{(t+1)}=(1-\beta_l)\delta^{(t)}$, with $\beta_l\in(0,1)$ as the decay rate. The discount factor $\zeta=0.99$ ensures higher cumulative rewards, benefiting V2V/V2N data transmission. The $\epsilon$-greedy algorithm starts with $\epsilon^{(0)}=0.1$ and adapts as $\epsilon^{(t+1)}=\max{0,(1-\beta_\epsilon)\epsilon^{(t)}}$, where $\beta_\epsilon=0.0001$. Batch size $D=100$ balances update noise and computational efficiency.
The DNNs used to implement the DQN consist of one input layer, three fully connected hidden layers, and one output layer. The input layer processes a state vector of length $|\mathbf{x}|=(2+K)\times N$, while the hidden layers have $H_1=5|\mathbf{x}|+8$, $H_2=3|\mathbf{x}|$, and $H_3=2|\mathbf{x}|$ neurons. This architectural design ingenuity reduces model size based on input feature reduction, balancing computational efficiency and accuracy. The rectifier linear unit (ReLU) and hyperbolic tangent (Tanh) functions are used alternatively as the activation function since this scheme leads to faster convergence, as per our observation.

\begin{table}[!t]
\caption{\textsc{Simulation parameters}}\footnotesize
\label{table_sim}
\centering
\scalebox{0.92}{
\begin{tabular}{|c|c|}
\hline
\textbf{Parameter} & \textbf{Value} \\
\hline
Carrier frequency & 2 GHz \\
\hline
Maximum transmit power $P_{\text{max}}$ & 23 dBm \\
\hline
Noise power $\sigma^{2}$ & -114 dBm \\
\hline
Vehicle receiver noise figure & 9 dB \\
\hline
Vehicle placement model & spatial Poisson \\
\hline
Carrier frequency $f_{c}$ & 2 GHz \\
\hline
Shadowing standard deviation $\sigma_{\text{sh}}$ & 3 dB \\
\hline
BS antenna height & 25 m \\
\hline
BS antenna gain & 8 dBi \\
\hline
BS receiver noise figure & 5 dB \\
\hline
Distance between RSU and highway & 35 m \\
\hline
Vehicle antenna height & 1.5 m \\
\hline
Vehicle antenna gain & 3 dBi \\
\hline
Payload size $L$ & 100 bits \\
\hline
Batch sampling period $\mathcal{T}_{b}$& 100 \\
\hline
Weight copy period $\mathcal{T}_{c}$ & 400\\
\hline
\end{tabular}
}
\end{table}

\begin{table}[!t]
\caption{\textsc{Hyperparameters for Deep-Q Networks}}
\label{table_DQN}
\centering
\scalebox{0.7}{
\begin{tabular}{|c|c|c|c|c|c|}
\hline
Layers & Input & layer 1 & layer 2& layer 3 & Output \\
\hline
Neuron number & $(2+K) \times N$ & 5$\|\mathbf{x}\|$+8 & 3$\|\mathbf{x}\|$ & 2$\|\mathbf{x}\|$ & $L_Q \times N$ \\
\hline
Activation function & Linear & Tanh & ReLU & Tanh & ReLU \\
\hline
Mini-batch size $D$ & \multicolumn{5}{c|}{100} \\
\hline
Optimizer & \multicolumn{5}{c|}{RMSProp} \\
 \hline
Discount factor $\zeta$ & \multicolumn{5}{c|}{0.99} \\
\hline
Initial learning rate $\delta^{(0)}$ & \multicolumn{5}{c|}{0.01} \\
\hline
Learning decay rate $\beta_l$ & \multicolumn{5}{c|}{0.0001} \\
\hline
Initial exploration rate $\epsilon^{(0)}$ & \multicolumn{5}{c|}{0.1} \\
\hline
$\epsilon$-decay rate $\beta_\epsilon$ & \multicolumn{5}{c|}{0.0001} \\
\hline
\end{tabular}
}
\end{table}
\begin{figure}[t]
    \centering
    \begin{subfigure}[t]{0.4\textwidth}
        \centering
        \includegraphics[width=\textwidth]{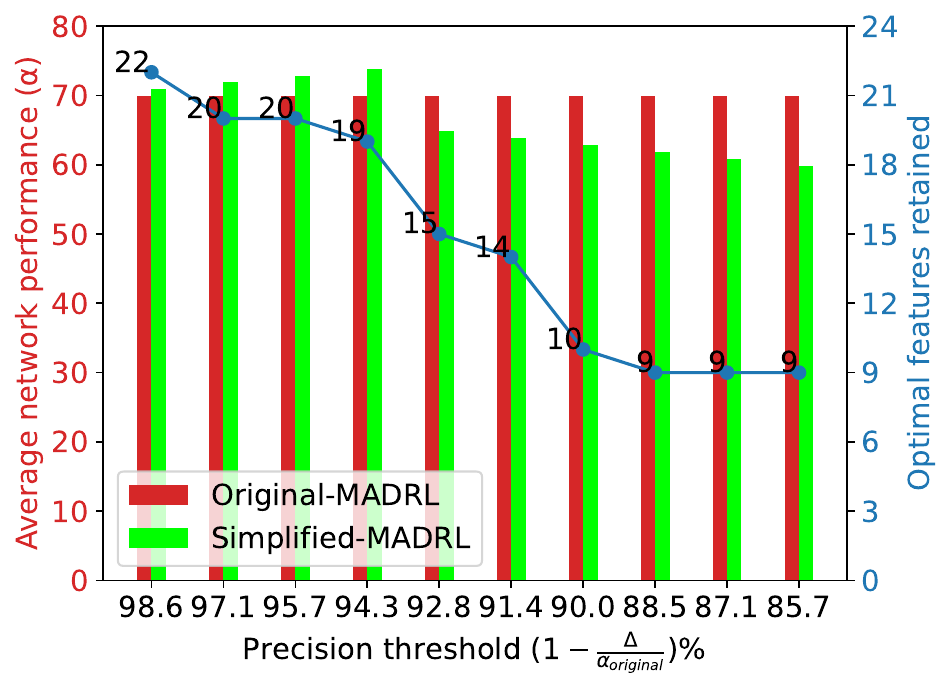} 
        \caption{}
        \label{ANR_}
    \end{subfigure}
    \hfill
    \begin{subfigure}[t]{0.4\textwidth}
        \centering
        \includegraphics[width=\textwidth]{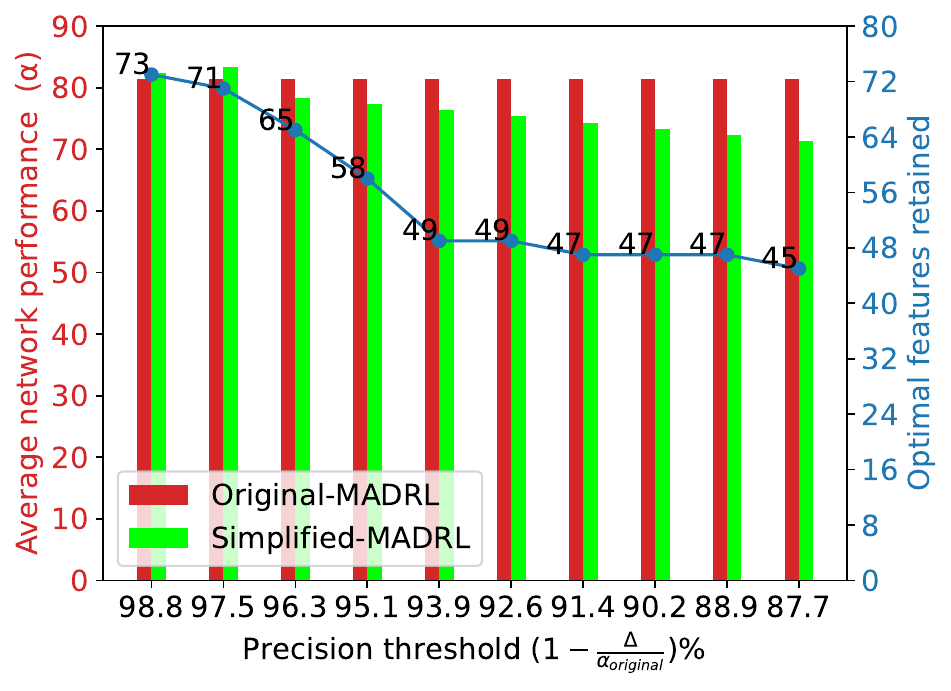}
        \caption{}
        \label{ANR_8}
    \end{subfigure}
    \caption{Average network performance and retained state features versus the precision threshold, with: (a)  $K=N=4$ where the Original-MADRL network has 24 state features per agent, and (b) $K=N=8$ where the Original-MADRL network has 80 state features per agent.}
    \label{ana}
\end{figure}

 \subsection{Analysis of XAI-based Variable State Feature Selection} \label{VFS} 
The main objective of our work is to simplify the proposed DRL-based framework through importance-based feature selection without losing performance. We first investigate the efficiency of the proposed importance-based variable feature selection methodology. Notice that the importance-based feature selection is a post-hoc process whereby the agents are first trained and tested using the Original-MADRL
 framework. Then, the summarized datasets $\mathcal{X}_{BG}$, $\mathcal{X}$  obtained from the training and execution process are utilized to obtain global state feature importance rankings using the systematic XAI-based methodology described in Section \ref{XAI_method}.

Fig. \ref{ANR_} shows the average network performance $\alpha$ and importance-based retained state features for different $\Delta$ values. Considering different QoS requirements for the V2V agents, we set different error probability constraints for each vehicle, $\varepsilon_{\text{max}, k} =$ \small$[10^{-4}, 10^{-3}, 10^{-4}, 10^{-3}]$\normalsize. It can be observed that an increase in $\Delta$ allows for retaining fewer optimal features. As seen, for $\Delta \leq 2$, the Simplified-MADRL network does not lose any performance compared to the Original-MADRL network. In fact, a slight performance improvement is observed due to the elimination of less important features that are most likely non-informative and add complexity to the network without providing performance benefits. For $\Delta >$ 2, there is a sharp decrease in the number of state elements, where approximately $38\%$ of the original state features are eliminated. This sharp decline is reflected by the decrease in $\alpha$, showing that some of the influential state input features have been removed.

Fig. \ref{ANR_8} shows the average network performance and retained state features for different $\Delta$ values in a network with  $N=8$ V2N links and $K=8$  V2V pairs. Here, we set  error probability constraints as $\varepsilon_{\text{max}, k} =$ \small$[10^{-3}, 10^{-4}, 10^{-2}, 10^{-4}, 10^{-3}, 10^{-2}, 10^{-3}, 10^{-4}]$\normalsize. Similar to the case for four V2V links, it can be observed that an increase in $\Delta$ results in fewer optimal state feature selections. For $\Delta =2$, the proposed variable feature selection methodology selects 58 optimal state features out of 80 for the Original-MADRL network while maintaining the average network performance gap less than $\sim$5\% between the Original-MADRL scheme and the Simplified-MADRL scheme, respectively. Based on the results from Fig. \ref{ANR_}, and
\ref{ANR_8}, there exists a trade-off between model input simplification (e.g., via feature selection) and network performance. To strike the right balance between feature selection and optimal performance in terms of the average network performance metric,  a moderate value of $\Delta =2$  is selected to achieve performance close to the Original-MADRL network while reducing the number of optimal state features by $\sim$21\% and $\sim$28\% for a network with $K$=4 and $K$=8 V2V pairs, respectively.
Therefore, for the subsequent simulation results, we fix the value of the precision threshold to $\Delta =2$. 

\subsection{Performance Comparison of Algorithms}

\begin{figure}[t]
    \centering
    \begin{subfigure}[t]{0.4\textwidth}
        \centering
        \includegraphics[width=\textwidth]{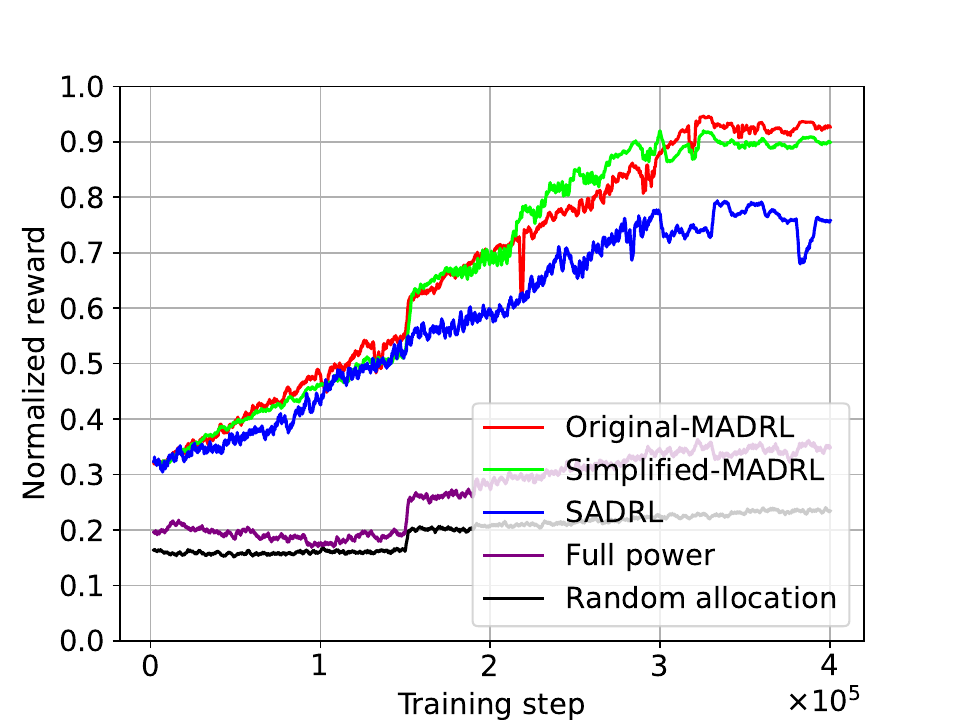}
        \caption{Training.}
        \label{learning4}
    \end{subfigure}
    \hfill
    \begin{subfigure}[t]{0.4\textwidth}
        \centering
        \includegraphics[width=\textwidth]{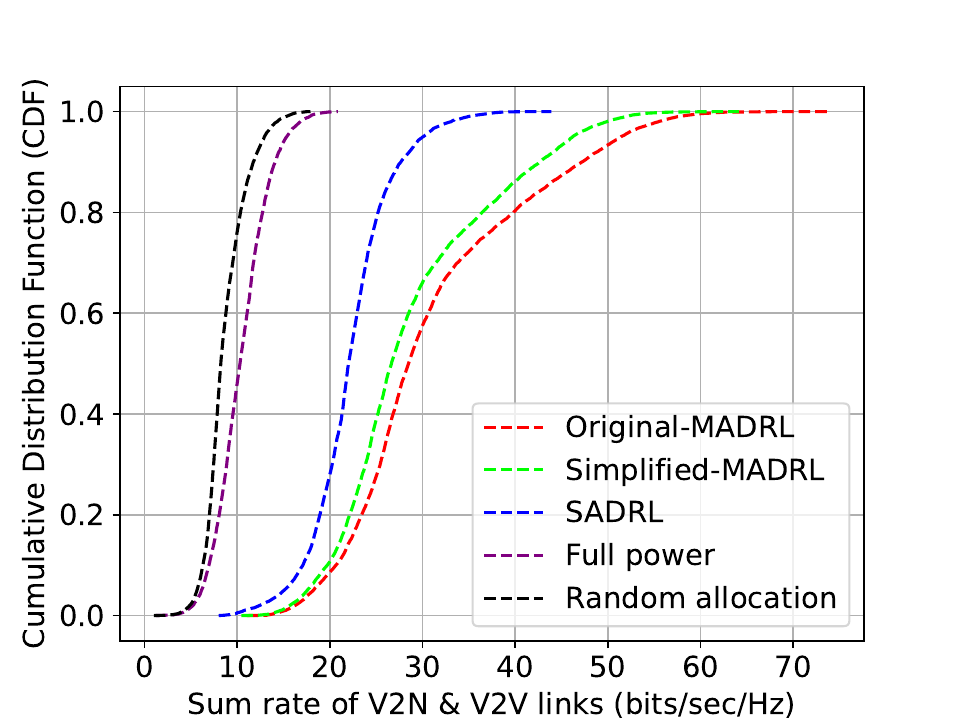}
        \caption{Testing- Empirical CDF.}
        \label{CDF4}
    \end{subfigure}
    \caption{Training and testing performance of network with $K = N = 4$, $\boldsymbol{\varepsilon}_{\max}=[\varepsilon_{\text{max}, 1}, \dots,\varepsilon_{\text{max}, k}, \dots, \varepsilon_{\text{max}, K}]=$\small$[10^{-4}, 10^{-3}, 10^{-4}, 10^{-3}]$\normalsize 
    ,  where (a) shows the learning process comparison for the different algorithms, (b) shows the empirical CDF of network sum-rate performance. Each value is the moving average of the previous 300 time slots.}
    \label{4vehicles}
\end{figure}

\subsubsection{Convergence Analysis and Sum rate performance}
Fig. \ref{learning4} illustrates the training convergence behavior of the different algorithms in terms of the reward performance in a network with $4$ V2V links and V2N links. It can be observed that the Original-MADRL scheme and the Simplified-MARL schemes converge around $3.2 \times 10^5$ and $2.8 \times 10^5$ time steps, respectively. In contrast, the centralized SADRL scheme demonstrates slow and suboptimal convergence, primarily due to the high complexity of its action space, where a single DQN determines power and sub-band allocation actions for all V2V links, rendering the approach impractical due to the significant transmission overhead involved. The Full power algorithm exhibits degraded performance as it always transmits at maximum power. Such resource allocation can lead to a lower sum rate and violation of reliability constraint if multiple links use the same sub-band. Moreover, the  Full power algorithm does not consider adapting the transmit power based on the link quality. The random allocation exhibits the worst performance because it randomly selects the sub-band and transmits power at each time step, irrespective of the link quality or available resources. The proposed XAI-based Simplified-MADRL scheme utilizing only 19 state features can remarkably converge to $\sim$97\% of the training performance of the Original-MADRL using 24 state feature inputs. This means that the proposed variable selection strategy can effectively filter the most relevant features contributing to the agents' actions without significant performance loss.

Fig. \ref{CDF4} shows the cumulative distribution function (CDF) of the achievable sum rate for the V2V and V2N links, comparing the performance of different algorithms during the deployment (inference) phase in a network with $K=N=4$ V2V and V2N links. Fig. \ref{CDF4} 
 further demonstrates the effectiveness
of the proposed  Simplified-MARL algorithm by empirically validating its performance against the different benchmark algorithms in the inference stage. The
 Simplified-MARL scheme  outperforms the SADRL scheme and
achieves on average $41\%$ higher achievable sum-rates for all the V2V and V2N links compared to the
SADRL scheme. Further, the average difference between the Simplified-MARL and the Original-MADRL schemes in terms of the sum-rate performance is less than $\sim$7\%.  This means that by
exploiting the well-trained DNNs, the proposed algorithm can effectively learn the patterns of the environment based on the optimal state features without requiring the full set of state feature inputs.

\begin{figure}[t]
    \centering
    \begin{subfigure}[t]{0.4\textwidth}
        \centering
        \includegraphics[width=\textwidth]{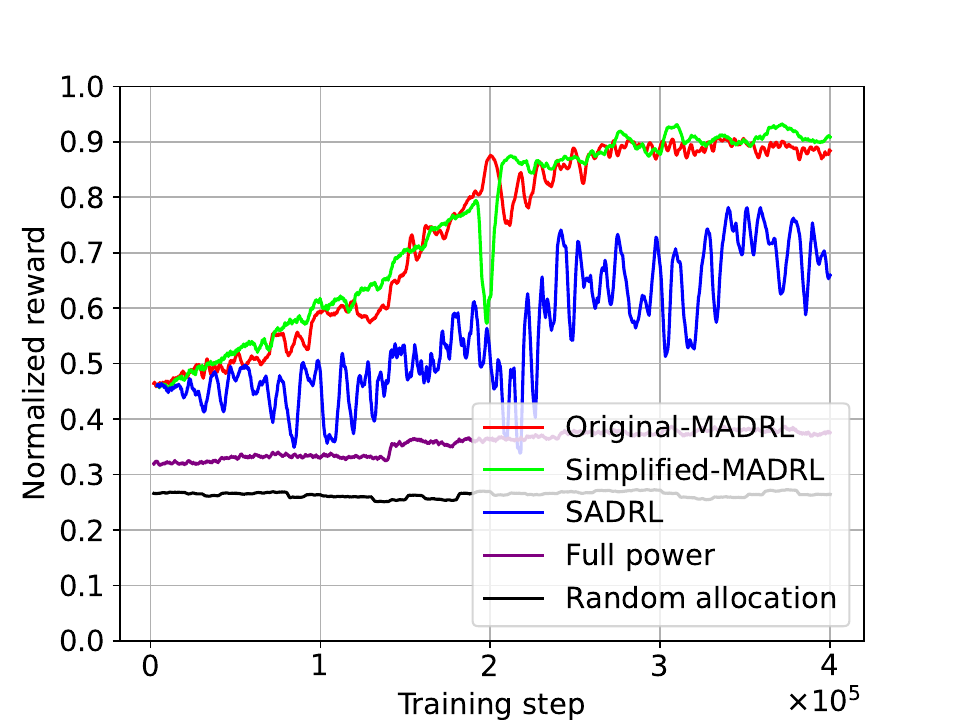}
        \caption{Training.}
        \label{learning8}
    \end{subfigure}
    \hfill
    \begin{subfigure}[t]{0.4\textwidth}
        \centering
        \includegraphics[width=\textwidth]{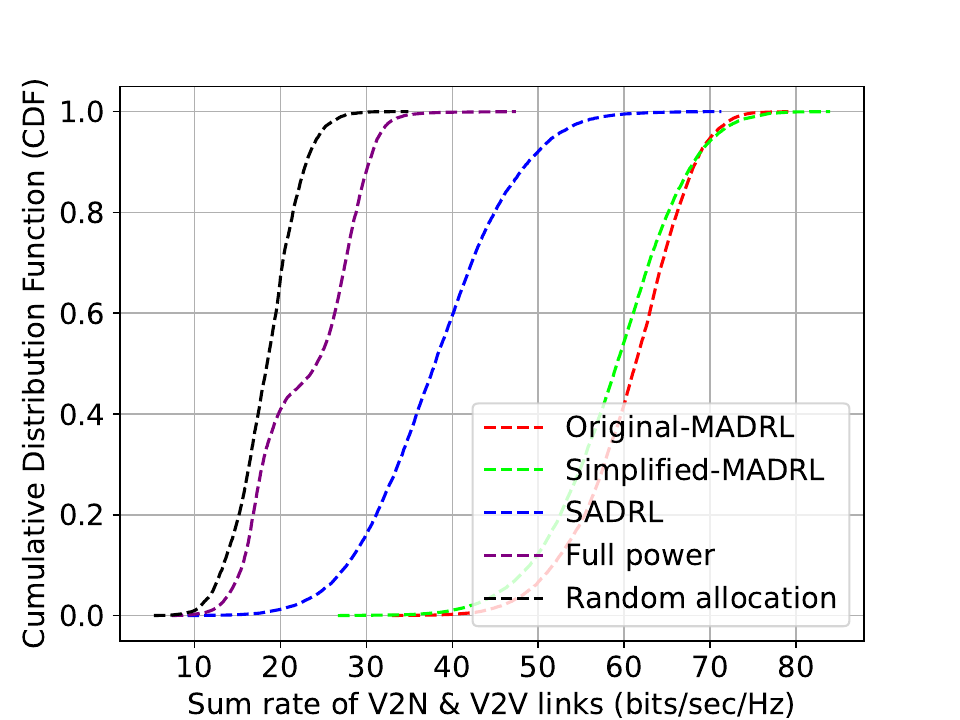}
        \caption{Testing- Empirical CDF.}
        \label{CDF8}
    \end{subfigure}
    \caption{Training and testing performance of network with  $K=N=$8, $\boldsymbol{\varepsilon}_{\max}=[\varepsilon_{\text{max}, 1}, \dots,\varepsilon_{\text{max}, k}, \dots, \varepsilon_{\text{max}, K}]=$ \small$[10^{-3}, 10^{-4}, 10^{-2}, 10^{-4}, 10^{-3}, 10^{-2}, 10^{-3}, 10^{-4}]$\normalsize, where (a) shows the learning process comparison for the different algorithms, (b) shows the Empirical CDF of network sum-rate performance. Each value is the moving average of the previous 300 time slots.}
    \label{8vehicles}
\vspace{-0.5cm}
\end{figure}

Fig. \ref{learning8} illustrates the training convergence behavior of different algorithms in a network with $K=N=$8 V2V and V2N links. It is observed that the rewards are low and unstable at the beginning of the learning as the training agents focus more on exploration than exploitation. Despite fluctuations due to mobility-induced channel fading in vehicular environments, the reward function increases and eventually converges with the increasing number of training iterations.
The Simplified-MADRL and the Original-MADRL schemes have comparable training performance, and both schemes converge in about $2.8 \times 10^5$ training steps. The SADRL scheme fails to achieve comparable performance to the proposed method and demonstrates reduced training efficiency under higher vehicular densities. This inadequacy is attributed to the rapid growth of the DQN network topology and action space as the vehicular network scales, making it challenging to thoroughly explore the entire action space to determine the optimal action selection.

 Similar to Fig. \ref{CDF4}, Fig. \ref{CDF8} shows the CDF of the achievable sum-rate for the V2V and V2N links; however,  in a network with $K=N=8$ V2V and V2N links.  The proposed approach outperforms the SADRL scheme, achieving, on average, a 66\% higher sum-rate in the testing stage. Interestingly, the Simplified-MADRL achieves higher sum-rates for the majority of the V2N and V2V links than the Original-MADRL scheme with $\sim$28\% fewer state features. This improvement is mainly attributed to eliminating less important and non-informative features that only complicate the learning process.

Note that compared to the recursive feature selection strategies requiring retraining for different subsets of features, the desired results for the Simplified-MARL algorithm are achieved after a single retraining of the models on the subset of important features predicted by Algorithm \ref{algo2}. As seen from Fig \ref{CDF4}, \ref{CDF8},  the simplified
agents have not lost significant performance when compared to the
original agents' performance. In fact,  a slight performance improvement in learning efficiency for larger network size is observed in Fig  \ref{8vehicles}  due to eliminating less important features.  This demonstrates that the proposed feature selection strategy effectively identifies the most relevant features for the multiple agent's actions with minimal impact on performance.

\begin{figure}[t]
    \centering
    \includegraphics[width=0.8\linewidth]{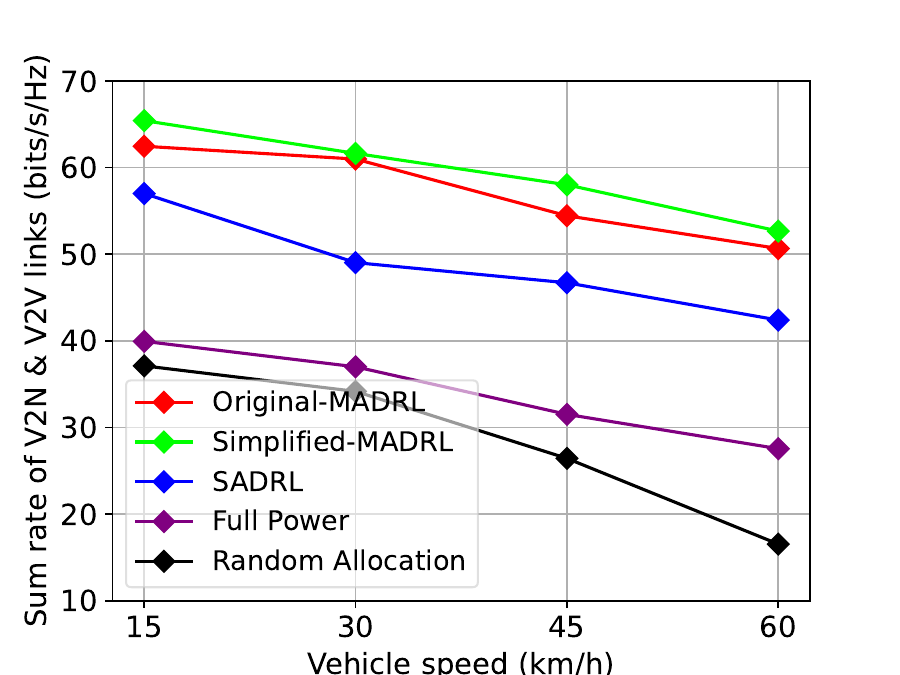}
    \caption{Sum rate performance of V2N and normal V2V links with varying
vehicle speeds.}
    \label{ratevsspeed}
\end{figure}

\subsubsection{Impact of Mobility}
Fig. \ref{ratevsspeed} shows the impact of vehicle speed on the sum data rate of V2N and V2V links, when the number of V2V and V2N links are set to $K=N=8$. From the figure, a general decline in sum rate performance is observed across all approaches as vehicle speed increases. This drop in performance is primarily due to reduced received power in V2X communication links, resulting from increased inter-vehicle distances at higher speeds, which lead to sparser traffic conditions. Additionally, higher speeds create observation uncertainties (e.g., channel state information and received interference), making it difficult to find an optimal policy, especially the random search approach, which shows the lowest performance. Our proposed Simplified-MADRL approach performs close to the Original-MADRL solution considering all state features during the learning process. Furthermore, the proposed MADRL-based solutions outperform the centralized SADRL and the Full power algorithms in terms of sum-rate performance across various vehicle speeds, adapting more effectively to dynamic networks.

\subsubsection{Reliability performance}
\begin{figure}[t]
    \centering
    \includegraphics[width=0.8\linewidth]{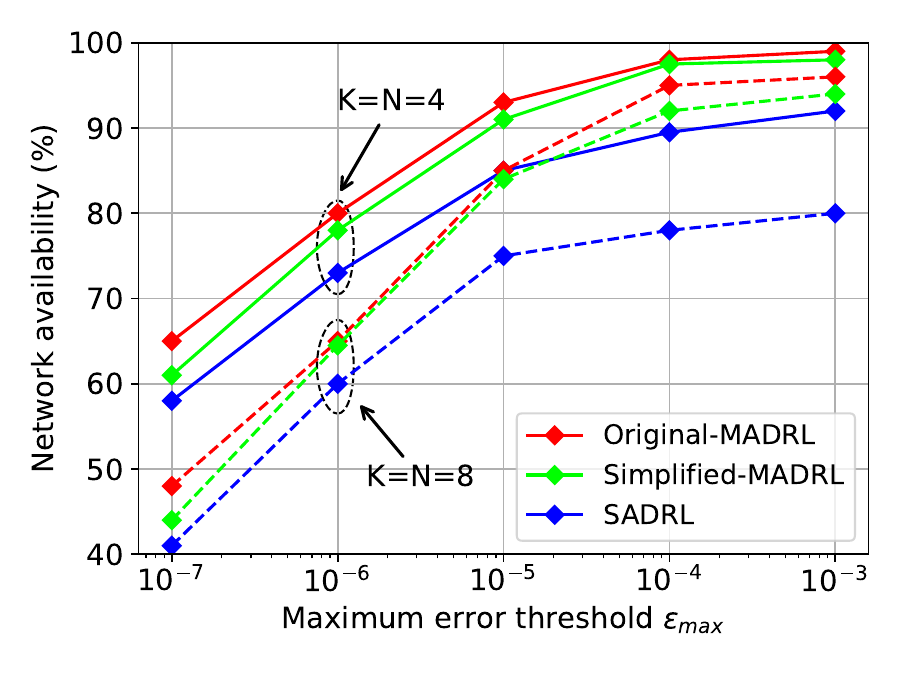}
    \caption{ Network availability for different maximum error probability thresholds.}
    \label{Link_reliability}
\end{figure}

In Fig. \ref{Link_reliability}, we consider the impact of reliability constraint violation. Thus, the performance metric is the network availability \cite {NA},
 defined as the percentage of the total number of
channel generations in the testing phase, for which the worst-case decoding error
probability, i.e., $\max\limits_{k \in K}\, \varepsilon^{(t)}_{k}[n]$ is no greater than a threshold \( \varepsilon_{\text{max}} \). Here, we keep the same maximum threshold error probability for all V2V links, i.e, \( \varepsilon_{\text{max}, k} \) = \( \varepsilon_{\text{max}} \) $\forall k \in K$. Across all schemes, the network availability percentage improves with increasing \( \varepsilon_{\text{max}} \) as the tolerable decoding error probability requirement becomes less stringent. Original-MADRL consistently achieves the highest reliability, reaching $\sim$92\% at \( \varepsilon_{\text{max}} = 10^{-5} \) for \( K=N=4 \) and nearly 98.3\% at  \( \varepsilon_{\text{max}} = 10^{-4} \). The proposed Simplified-MADRL follows closely, with only a slight gap below Original-MADRL, achieving around 98\% at  \( \varepsilon_{\text{max}} = 10^{-4} \) for \( K=N=4 \). Network availability decreases as the number of V2V and V2N links increases due to fixed resource sharing and higher interference, worsening reliability violations. Simplified-MADRL maintains reasonable reliability for \( \varepsilon_{\text{max}} > 10^{-5} \), performing close to Original-MADRL. In contrast, centralized SADRL falls behind, achieving only \(\sim 80\%\) for \( K = N = 8 \) at \( \varepsilon_{\text{max}} = 10^{-3} \). The gap is largest at lower error levels, highlighting the resilience of Original-MADRL and Simplified-MADRL to increased V2V link density.

\subsubsection{Scalability}
\begin{figure}[t]
    \centering
    \includegraphics[width=0.8\linewidth]{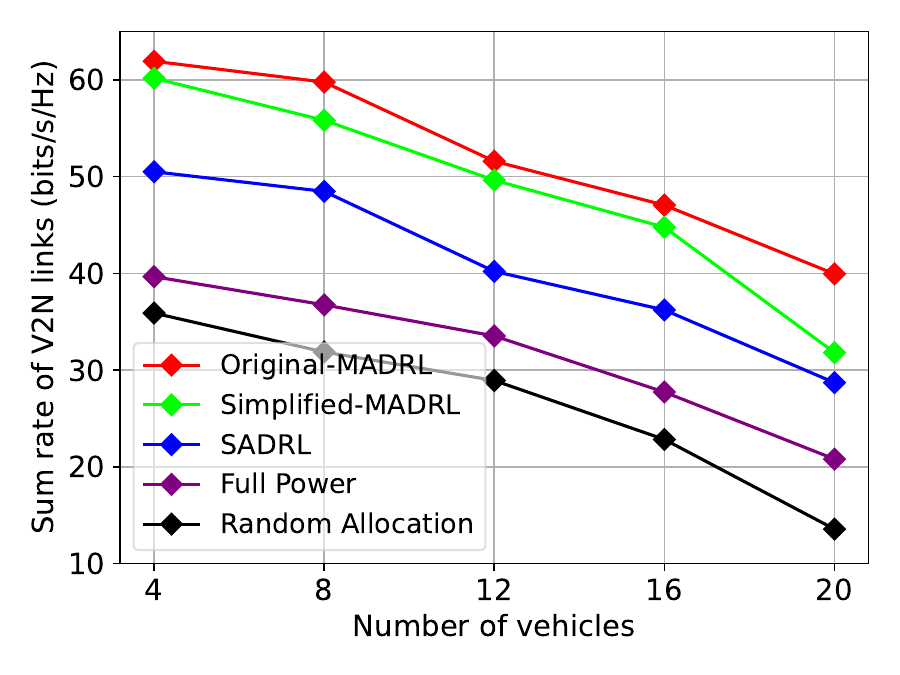}
    \caption{ Sum rate  of V2N links  with varying
number of vehicles.}
    \label{ratevsvehicles}
\end{figure}

Fig. \ref{ratevsvehicles}  demonstrates the impact of varying number of vehicles on the sum rate of the V2N links. Here, we fix the number of sub-bands for V2N links to $N=4$ and increase the number of V2V links to assess the impact of dynamic vehicle numbers on the sum-rate performance. From the figure, it is observed that when
  the number of vehicles increases, more V2V links share the
fixed number of sub-bands, which causes stronger interference to
the V2N links, resulting in a rate reduction for all schemes. The proposed Simplified-MADRL approach achieves better performance than the other baseline schemes, demonstrating its robustness across different numbers of vehicles. The centralized SADRL approach shows degraded performance mainly
caused by the increased size of the joint learning action
space and increased DQN output layer complexity, whereas the poor performance of the Full power algorithm can be explained by the strong interference due to maximum power transmission, negatively impacting the sum rate of the V2N links.

\subsection{Computational Complexity Reduction Analysis}
The computational complexity of the proposed learning algorithms depends on the structure of implemented DNNs (number of hidden layers and the number of corresponding neurons). The complexity of learning algorithms is primarily dominated by the update process of the network weight parameters. Each V2V agent requires a periodic update of its local DQN parameters during training. The input layer of each DQN always matches the number of input features, and the number of neurons in the hidden layers depends on these input features. Therefore, reducing the number of input features decreases the computational load for propagating information through the network and also reduces the number of parameters that need to be periodically broadcasted to update the local agents' DQNs.

The total number of scalar parameters to broadcast per agent is $ |\mathbf{\theta}|= \left(\|\mathbf{x}\| + \sum_{l=1}^3H_l H_{l+1}\right)$. The update process of the loss function is linear to the mini-batch size $D$ since the optimizer scans all mini-batch samples. As a result, an update of all parameters incurs a computational complexity of $\mathcal{O}\left(D  (\|\mathbf{x}\| + \sum_{l=1}^3H_l H_{l+1}) \right)$. Considering the proposed Simplified-MADRL approach, the overall computational complexity
is $\mathcal{O}\left(KD (\|\mathbf{x}\| + \sum_{l=1}^3H_l H_{l+1}) \right)$ for updating the network parameters of all
$K$ agents. Table (\ref{table_time})  shows the complexity comparison in terms of the average training time, the number of parameters to broadcast, and the number of state features used according to the variable feature selection methodology. Interestingly, the simplified-MADRL reduces average training time computed across training steps by  $\sim$26\%  and $\sim$11\% for $K$=4 and $K$=8 V2V agents, respectively. In terms of the the broadcast parameters per agent, the proposed approach achieves a reduction of  $\sim$35\% and $\sim$46\% by utilizing 19 and 58 most important state features, compared to the 24 and 80 state features used by the Original-MARL agent. In our evaluation environment, the agents spend an average of $8.3 \times 10^{-4}$ seconds for each action selection. There is considerable potential for optimizing execution efficiency since the computational capabilities of processing units in real-world V2X networks exceed those of the computers utilized in     lations. 

\captionsetup{font=small, justification=centering}
\begin{table}[t!]
    \centering
    \caption{\textsc{Complexity comparison of algorithms.}}
    \label{table_time}
    \small
    \renewcommand{\arraystretch}{1.2} 
    \scalebox{0.75}{
    \begin{tabular}{|c|c|c|c|c|c|c|}
    \hline
    \textbf{Scheme} & \multicolumn{2}{c|}{\makecell{\textbf{Average Training}\\ \textbf{Time} (\si{\milli\second})}} & \multicolumn{2}{c|}{\makecell{\textbf{Number of}\\ \textbf{broadcast parameters}}} & \multicolumn{2}{c|}{\makecell{\textbf{Number of}\\ \textbf{state features}}} \\
    \hline
    & {K=4} & {K=8} & {K=4} & {K=8} & {K=4} & {K=8} \\
    \hline
    Original-MADRL & 6.23 & 13.4 & 34,544  & 347,920 & 24& 80\\
    \hline
    Simplified-MADRL & 4.56 & 11.9 & 22,424 & 185,912 & 19 & 58 \\
    \hline
    \end{tabular}}
\end{table}

\subsection{Discussion on Implementation and Future Extensions:}
\textit{Practical implications:} The proposed MADRL-based solution for spectrum and power resource allocation requires an offline training phase and a deployment phase. Offline training is employed to mitigate the computational overhead of learning the mapping between input states and output actions. Subsequently, during online network deployment, this mapping is executed using the pre-trained DNNs, bypassing the need for intensive computations. Offline training can be performed using a digital twin of the V2X network, incorporating network topology, channel
models, and QoS requirements. The proposed 
algorithm can utilize this twin for offline initialization
and network training at the BS to explore the simulated environment, mitigating real-world risk. Further, in dynamic vehicular networks, it is challenging to provide theoretical convergence guarantees for the proposed MADRL-based solution and other advanced multi-agent adaptations, despite their good empirical performances \cite{MADRL_review1}. 
By giving all agents a common global reward, the proposed MADRL-based solution mitigates the instability
of the multi-agent environment but makes each agent fail to achieve
the higher individual reward. The proposed XAI-based framework could be easily
adapted to work with other multi-agent frameworks and cooperative value decomposition-based methods such as QMIX \cite{Qmix_journal} and its adaptations to improve convergence stability and cooperation among the agents.

\textit{Dynamic network topology:} It is worth pointing
out that the well-trained DRL models
can be easily outdated due to the high mobility of vehicles. In our proposed centralized training and decentralized implementation approach, the weight parameters of the DNNs are centrally trained
at the BS. The BS periodically broadcasts the latest trained weight parameters to the local
agents. Retraining can be triggered if the
network performance falls below a threshold.
Further, considering nearby V2V pairs often experience similar
channel quality and environment observations, the RSU can intelligently cluster vehicles into a specific zone and use the zone information to train newly activated
V2V pairs. For newly activated V2V pairs, they request the BS for connectivity. The RSU has complete information on the vehicles' location and network topology. For an incoming vehicle, the RSU establishes the corresponding DNNs and transfers the weights of the already-trained model to the newly established DNNs, which are downloaded upon request from the incoming vehicle. To accommodate the changing number of V2V agents, state encoding adjusts for changes in the number of V2V agents under RSU coverage.  If the number of vehicles exceeds the maximum number of trained agents, information from the extra vehicles is not included
in the state. Moreover, if one or more vehicles leave the coverage area, the RSU can introduce virtual agents with their direct and interfering channel gains set to zero to avoid impacting other agents' decision-making. The purpose of having these virtual agents as placeholders is
to provide inconsequential inputs to fill the input elements
of fixed length, i.e., the agent’s state is padded with zeros. Further, the proposed XAI-based feature selection strategy somehow mitigates this issue since all trained agents use a fixed size of importance-based state features during training and testing.

\textit{Generalization to different environments:} The proposed method is designed to adapt to the specific characteristics of a given vehicular environment. However, when the environment changes, the DNNs need to be retrained, and the feature selection needs to be reevaluated. In such cases, the
multiple agents can be trained from scratch or initialized
with the existing DNN weights, then refined using data from the new environment. To enhance the adaptability of the proposed MADRL-based solution to new environment scenarios and tasks
requiring computation-intensive training, model-agnostic meta-learning \cite{MetaV2X, Meta} can be incorporated into
the proposed framework.

\section{Conclusions}\label{sec:conclusion}
In this paper, we have developed a model-agnostic XAI-based methodology to explain the inference process of multiple trained DRL agents and have used it to simplify the DRL agents' input state size through importance-based feature selection. We have capitalized on SHAP  for generating feature importance explanations and reducing the complexity of multiple trained  models by generating a new model with a reduced size that requires lower training time and reduced network parameters to broadcast while maintaining a reasonable performance. We have applied the proposed methodology to a practical vehicular communication network where multiple agents are first trained using a novel centralized learning and decentralized execution algorithm. Subsequently, the XAI-based framework is applied to the trained models to select important features using an automated feature selection strategy. Our results have shown that the XAI-assisted methodology remarkably reduces the optimal state features by  $\sim$21\% and $\sim$28\%, average training time by  $\sim$26\% and $\sim$11\%,  and trainable weight parameters to broadcast by $\sim$35\% and $\sim$46\% in a network setting with four and eight vehicular pairs, respectively. In the future, we plan to adapt the proposed XAI-based methodology for other wireless resource allocation problems. Additionally, applying XAI to digital twin systems to simplify complex virtual representations of real-world networks and maintain synchronization between the twin and the actual network is a promising research direction. By interpreting model actions using environment semantics, XAI can enhance the connection between digital twins and real-world dynamics.

\bibliographystyle{IEEEtran}\bibliography{IEEE-abr,XAI-bibliography}

\end{document}